\RequirePackage{fix-cm}
\documentclass[smallextended]{svjour3}    
\smartqed  
\usepackage{adjustbox}
\usepackage{multirow}
\usepackage{appendix}
\usepackage{booktabs,caption}
\usepackage[flushleft]{threeparttable}
\usepackage{orcidlink}
\usepackage{comment}
\usepackage {drafts}
\usepackage{subfigure}
\usepackage{listings}
\usepackage{courier}
\usepackage{array}
\usepackage{longtable}
\usepackage{circledsteps}
\usepackage{float}

\usepackage{xurl}

\usepackage{cite}
\usepackage{amsmath,amssymb,amsfonts}
\usepackage{algorithmic}
\usepackage{graphicx}
\usepackage{textcomp}
\usepackage{xcolor}
\usepackage{url}
\usepackage[nolist]{acronym}
\usepackage{xspace}
\usepackage{tikz}

\newcommand*\circled[1]{%
    \tikz[baseline=(char.base)]{%
        \node[shape=circle,draw={rgb,255:red,0;green,31;blue,38},%
              fill={rgb,255:red,0;green,77;blue,104},%
              text=white,inner sep=.3pt] (char) {{{\texttt\textbf \small#1}}};%
    }%
}

\newcommand*{\ie}{i.e.,\@\xspace}

\newcommand*{\jjodel}{{Jjodel\@\xspace}}
\makeatletter
\newcommand*{\etc}{%
	\@ifnextchar{.}%
	{etc}%
	{etc.\@\xspace}%
}
\makeatother

\newcommand{\code}[1]{{\texttt{#1}}}

\definecolor{codegray}{rgb}{0.5,0.5,0.5}
\definecolor{attributes}{rgb}{0.5,0.5,0.5}
\definecolor{codepurple}{rgb}{0.58,0,0.82}
\definecolor{backcolour}{rgb}{0.95,0.95,0.92}

\definecolor{codered}{HTML}{880044}
\definecolor{codegreen}{HTML}{129490}
\definecolor{codeblue}{HTML}{0A2E36}

\lstdefinestyle{mystyle}{
    backgroundcolor=\color{backcolour},   
    commentstyle=\color{codegreen},
    keywordstyle=\color{codered},
    numberstyle=\tiny\color{codegray},
    stringstyle=\color{codeblue},
    basicstyle=\ttfamily\footnotesize\tiny,
    breakatwhitespace=false,         
    breaklines=true,                 
    captionpos=b,                    
    keepspaces=true,                 
    numbers=left,                    
    numbersep=5pt,                  
    showspaces=false,                
    showstringspaces=false,
    showtabs=false,                  
    tabsize=2,
    keywords={data,value},keywordstyle=\color{codered},
    morekeywords={val,left,right},keywordstyle=\color{codegreen}
}

\definecolor{codered}{HTML}{880044}
\definecolor{codeazure}{HTML}{129490}
\definecolor{codeblue}{HTML}{2660A4}

\lstdefinelanguage{Jjodel}
{
    keywords=[1]{value , DObject, self, instanceof , name},
    keywordstyle=[1]\color{codered},
    keywords=[2]{\$val ,\$left ,\$right },
    keywordstyle=[2]\color{codeazure},
    keywords=[3]{data ,node, view, context, inv},
    keywordstyle=[3]\color{codeblue},
    sensitive=false,
    morestring=[b]',
    morecomment=[l]{--}
}

\usepackage{inconsolata}

\makeatletter
\newcommand{\srcsize}{\@setfontsize{\srcsize}{7pt}{8pt}}
\makeatother
\xdefinecolor{green}{RGB}{63,128,35}
\xdefinecolor{plum}{RGB}{143,0,0}
\xdefinecolor{gray}{RGB}{153,153,153}
\xdefinecolor{red}{RGB}{204,77,101}
\xdefinecolor{blue}{RGB}{51,117,166}
\xdefinecolor{dblue}{RGB}{50,50,200}
\xdefinecolor{commentgreen}{RGB}{30,140,100}

\xdefinecolor{eminence}{RGB}{108,48,130}
\xdefinecolor{weborange}{RGB}{255,165,0}
\xdefinecolor{frenchplum}{RGB}{129,20,83}

\lstdefinelanguage{jsx}{
    basicstyle=\fontfamily{pcr}\srcsize,
    xleftmargin=0pt,
    alsodigit = {-},
    comment=[l]{//},
    morecomment=[s]{/*}{*/},
    commentstyle=\color{commentgreen},
    stringstyle=\color{weborange},
    keywords = {data,node,view},
    keywordstyle=\color{weborange},
    classoffset=1,
    otherkeywords={>,<,.,;,-,!,=,~},
    morekeywords={>,<,.,;,-,!,=,~},
    keywordstyle=\color{plum},
    classoffset=2,
    keywordstyle=\color{plum},
    otherkeywords = {values, oclCondition, name, x, y},
    morekeywords = {values, oclCondition, name, x, y},
    classoffset=3,
    keywordstyle=\color{blue},
    otherkeywords = {ownedAttributes},
    morekeywords = {ownedAttributes},
    classoffset=0
}

\lstdefinelanguage{css}{
    basicstyle=\fontfamily{pcr}\srcsize,
    morecomment=[s]{/*}{*/},
    commentstyle=\color{commentgreen},
    xleftmargin=0pt,
    alsodigit = {-},
    keywords = {grid},
    keywordstyle=\color{green},
    classoffset=1,
    otherkeywords={>,<,.,;,:,!,=,~},
    morekeywords={>,<,.,;,:,!,=,~},
    keywordstyle=\color{gray},
    classoffset=2,
    keywordstyle=\color{plum},
    otherkeywords={15,background-image,background-size,background-position},
    morekeywords={15,background-image,background-size,background-position},
    classoffset=3,
    keywordstyle=\color{red},
    otherkeywords={radial-gradient},
    morekeywords={radial-gradient},
    classoffset=4,
    keywordstyle=\color{plum},
    otherkeywords={0,1,2,3,4,5,6,7,8,9},
    morekeywords={0,1,2,3,4,5,6,7,8,9},
}

\lstdefinelanguage{JSX-template}{
    basicstyle=\ttfamily\srcsize,
    comment=[l]{//},
    morecomment=[s]{/*}{*/},
    commentstyle=\color{commentgreen},
    morekeywords = [1]{View, Control, Slider, Input, Selector, Edge, DefaultNode, decorators, values, value, id, node},
    morekeywords = [2]{true, false, div, map},
    morekeywords = [3]{name, ownedAttributes, isPK, left, right},
    morekeywords = [4]{className, field, hidden, autosize, style, start, end, key, data, view, title, payoff, min, max},
    keywordstyle= [1]\color{red},
    keywordstyle = [2]\color{plum},
    keywordstyle = [3]\color{dblue},
    keywordstyle = [4]\color{green},
    sensitive = true,
    morestring = [s]{="}{"},
    morestring = [b]',
    stringstyle = \color{blue},
    literate=%
        {\{data}{\{{\color{red}data\color{black}}}5
        {\$}{{\textcolor{dblue}{\$}}}1%
        {>}{{\textcolor{plum}{>}}}1%
        {=>}{{\textcolor{plum}{=>}}}2%
        {>=}{{\textcolor{plum}{>=}}}2%
        {===}{{\textcolor{plum}{===}}}3%
        {!==}{{\textcolor{plum}{!==}}}3%
        {\&\&}{{\textcolor{plum}{\&\&}}}2%
        {<}{{\textcolor{plum}{<}}}1%
        {/>}{{\textcolor{plum}{/>}}}2%
        {</}{{\textcolor{plum}{</}}}2%
        {=\{}{{\textcolor{plum}{=\{}}}2%
        {\}}{{\textcolor{plum}{\}}}}1%
        {\{}{{\textcolor{plum}{\{\kern-0pt}}}1%
        {)\}}{{\textcolor{plum}{)\}}}}1%
        {)}{{\textcolor{plum}{)}}}1%
        {(}{{\textcolor{plum}{(}}}1%
        {\{data}{{\textcolor{plum}{\{\textcolor{red}{data}}}}5%
        {=\{data}{{\textcolor{plum}{=\{\textcolor{red}{data}}}}5%
}

\def\BibTeX{{\rm B\kern-.05em{\sc i\kern-.025em b}\kern-.08em
    T\kern-.1667em\lower.7ex\hbox{E}\kern-.125emX}}
\begin{document}

\title{Modeling in \jjodel: Bridging Complexity and Usability in Model-Driven Engineering}

\titlerunning{Modeling in \jjodel{}}        

\author{Antonio Bucchiarone \orcidlink{0000-0003-1154-1382}        \and
        Juri Di Rocco\orcidlink{0000-0002-7909-3902}     \and 
        Damiano Di Vincenzo \and
        Alfonso Pierantonio  \orcidlink{0000-0002-5231-3952} 
}

\institute{SWEN, Università degli Studi dell'Aquila, 67100, L'Aquila, Italy\\ \email{\{name.surname\}@univaq.it}\\ \email{damiano.divincenzo@studenti.univaq.it}
}

\authorrunning{A. Bucchiarone et al.} 

\date{Received: date / Accepted: date}

\maketitle
\begin{abstract}

\jjodel{} is a cloud-based reflective platform designed to address the challenges of Model-Driven Engineering (MDE), particularly the cognitive complexity and usability barriers often encountered in existing model-driven tools. This article presents the motivation and requirements behind the design of \jjodel{} and demonstrates how it satisfies these through its key features. By offering a low-code environment with modular viewpoints for syntax, validation, and semantics, \jjodel{} empowers language designers to define and refine domain-specific languages (DSLs) with ease. Its innovative capabilities, such as real-time collaboration, live co-evolution support, and syntax customization, ensure adaptability and scalability for academic and industrial contexts. A practical case study of an algebraic expression language highlights the ability of \jjodel{} to manage positional semantics and event-driven workflows, illustrating its effectiveness in simplifying complex modeling scenarios. Built on modern front-end technologies, \jjodel{} bridges the gap between theoretical MDE research and practical application, providing a versatile and accessible solution for diverse modeling needs.

\end{abstract}

\keywords{Model-Driven Engineering (MDE) \and reflective platforms \and low-code development \and \jjodel{} \and collaborative modeling \and co-evolution}

\section{Introduction}\label{sec1}
The growing complexity of software systems demands tools that allow for efficient and intuitive modeling~\cite{abrahao2017user}. MDE has emerged as a powerful paradigm to facilitate the transition from abstract models to concrete implementations \cite{3103551,10350721}. However, many existing modeling tools inadvertently add complexity~\cite{brooks1987essence}, especially in educational settings where the primary objective is to simplify learning and make modeling more accessible to students and newcomers \cite{Liebel2017}.

Modeling tools play a critical role in software development, offering designers and engineers a means to explore design spaces and communicate complex systems effectively to stakeholders~\cite{Kienzle2024}. Despite their importance, academic tools often fail to meet expectations, lacking maturity, robustness, and usability. Students are frequently confronted with issues such as cumbersome installation, configuration, and maintenance processes, which shift their focus from understanding modeling concepts to overcoming technical obstacles. In contrast, while industrial tools tend to be more polished and feature rich, they are often prohibitively expensive, difficult to learn, and misaligned with the specific needs of educational settings~\cite{Kienzle2024}.  

These challenges highlight the need for tools that simplify the modeling process without compromising capability. An ideal tool for education should be intuitive, easy to use and platform independent, while supporting various paradigms and DSLs~\cite{Kienzle2024}. Furthermore, modern tools must showcase the practical benefits of MDE by allowing students to create models that drive downstream development activities, rather than serving as static diagrams~\cite{Stahl2006}. 

\smallskip
This article introduces \jjodel{}~\cite{di2021enhancing,di2023jjodel}\footnote{The \jjodel{} tool available online at \url{https://www.jjodel.io}. The source code can be accessed on GitHub at \url{https://github.com/MDEGroup/jjodel}. Furthermore, a set of video tutorials demonstrating various features and use cases of \jjodel{} are available at \url{https://www.jjodel.io/video-tutorials/}.}, a cloud-based reflective platform designed to address these challenges. Through a practical example, we demonstrate how \jjodel{} streamlines the modeling process by offering a low-code environment that balances simplicity and advanced functionality, enabling diverse modeling tasks while prioritizing accessibility and usability.  
Building on its emphasis on \textit{transparency in tools}\footnote{The idea of transparency in tools has been developed and explored by several scholars, particularly in the domains of philosophy, technology studies, and human-computer interaction (HCI). Although no single individual is solely credited with the concept, a key contribution has come from Martin Heidegger with his work \textit{Being and Time} (1927)~\cite{heidegger1927being}.}, \jjodel{} seeks to eliminate technical barriers that often disrupt the modeling process, ensuring that modelers can remain focused on their objectives rather than struggling with tool operation. By prioritizing accessibility for users of varying level of expertise, \jjodel{} integrates advanced features such as syntax viewpoints, real-time validation, and collaborative modeling, offering a seamless and efficient environment for diverse modeling tasks. 

Unlike many academic tools, \jjodel{} removes the challenges of installation and configuration, allowing modelers to focus on modeling concepts and their practical applications. Its features, including real-time collaboration, automated feedback, and support for multiple modeling paradigms, bridge the gap between the simplicity required for modeling and the robustness demanded by software development. By making modeling more accessible and empowering, \jjodel{} provides an intuitive platform that not only enhances learning, but also demonstrates the real-world value of MDE.

In this article, we demonstrate how \jjodel{} transforms the language engineering process. Its innovative approach simplifies the learning curve, offering students an opportunity to focus on mastering MDE principles and applying them effectively in diverse scenarios. Through its integration of accessibility, functionality, and educational support, \jjodel{} emerges as a compelling solution to the challenges of teaching modeling in modern educational environments.

\subsection{Objectives and Structure the Article}

This article introduces \jjodel{}, a cloud-based platform that tackles the challenges of MDE by combining accessibility, advanced features, and educational support. By simplifying complexity and enabling real-time collaboration, automated co-evolution, and modular syntax viewpoints, \jjodel{} connects more theoretical research with practical MDE applications.


The discussion opens with an in-depth overview of \jjodel{}’s architecture and key features (Section \ref{sec:overview}), detailing its front-end and back-end components, object model, and dynamic viewpoints that collectively enhance the modeling process. To demonstrate its practical utility, a case study of an algebraic expression language is presented (Section \ref{sec:practice}), illustrating \jjodel{}’s capabilities in managing abstract and concrete syntax, validation, and event-driven workflows. 
The article reflects on the lessons learned during the development of \jjodel{}, emphasizing its broader implications for the MDE community (Section \ref{sec:lessons}).
Section \ref{sec:rw} reviews related tools and highlights how \jjodel{} addresses specific challenges in MDE. Conclusions outline future directions (Section \ref{sec:conclusion}), with a focus on strategies to improve scalability, integration, and community participation. Furthermore, it highlights the opportunity to showcase a demonstrator, further solidifying \jjodel{}’s position as a versatile and impactful tool for addressing diverse MDE challenges.

\section{Overview of \jjodel{}}
\label{sec:overview}

\begin{figure}[t]
    \centering
    \subfigure[ERD metamodel]{
        \includegraphics[width=.44\textwidth]{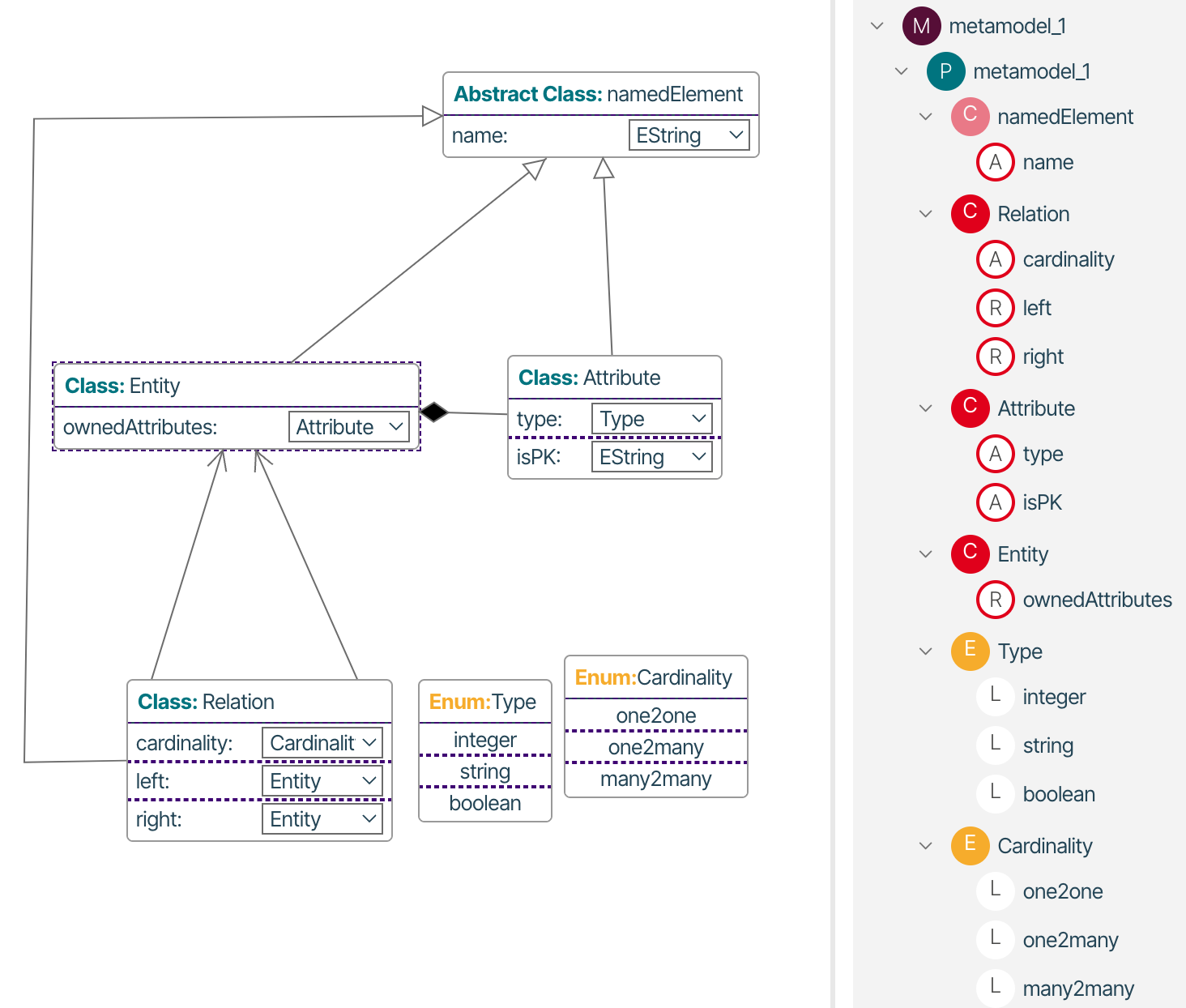} 
    }
    \subfigure[ERD instance]{
        \includegraphics[width=0.51\linewidth]{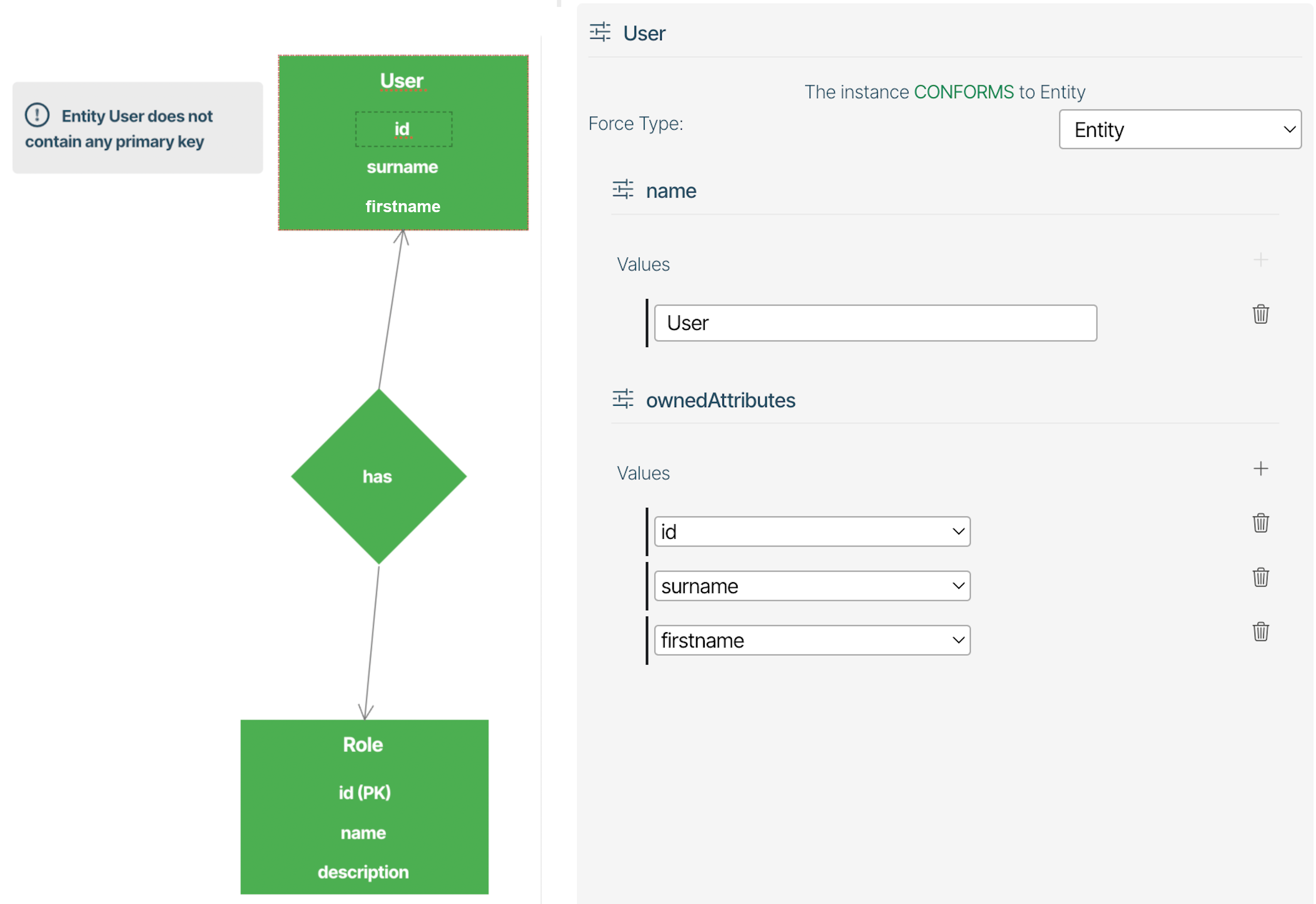}
    }\\[5mm]
    \caption{Entity Relationship Diagram in \jjodel{}.}
    \label{fig:erd}
\end{figure}

To provide a comprehensive understanding of \jjodel{}’s capabilities, this section explores its architecture and core components. Section~\ref{sec:architecture} begins with an overview of the general structure of \jjodel{}, followed by an in-depth analysis of its front- and back-end services, the \jjodel{} Object Model (JjOM), the syntax and validation viewpoints, and the editing mechanisms. Together, these components illustrate how \jjodel{} enhances MDE workflows by offering flexibility, supporting unobtrusive co-evolution, enabling seamless collaboration, and ensuring robust and efficient functionality.

We illustrate these features using an example of an entity relationship diagram (ERD). In particular:

\begin{itemize}
    \item Figure~\ref{fig:erd}(a) illustrates the metamodel of an ERD, defined using classes and enumerations. An \code{Entity} contains \code{owned\-Attributes}, where each \code{Attribute} is characterized by a \texttt{type} and an \texttt{isPK}~attribute, indicating whether it is a primary key. A \texttt{Relation} connects two entities through the \texttt{left} and \texttt{right} references and is constrained by a \texttt{Cardinality} specification.
    
    \item Figure~\ref{fig:erd}(b) shows an example of an ERD model that conforms to the metamodel and is represented with a concrete syntax representing a diagrammatic notation. For example, the \texttt{User} entity includes attributes such as \texttt{id}, \texttt{surname}, and \texttt{firstname}, while the \texttt{Role} entity includes \texttt{id}, \texttt{name}, and \texttt{description}. These entities are connected by the \texttt{has} relationship, illustrating how the model adheres to the defined metamodel.
\end{itemize}

This example highlights the ability of \jjodel{} to handle both metamodel definitions and conforming models in an integrated environment, ensuring flexibility and precision in MDE workflows. We continue this section by presenting the architecture of \jjodel{}, detailing its core components and how they collectively support efficient MDE practices.

\subsection{\jjodel{} Architecture}\label{sec:architecture}
\jjodel{} is a cloud-based reflective modeling platform that enhances MDE by minimizing accidental complexity and simplifying workflows. It offers robust capabilities, including metamodel and model definition, concrete syntax specification, and model validation using user-defined constraints. Moreover, with support for collaborative modeling in real-time, \jjodel{} enables multiple users to work simultaneously on shared projects, fostering teamwork and efficiency.


The architecture of \jjodel{} is built on a carefully selected technology stack that not only supports real-time collaboration, efficient workflows, and robust back-end functionality, but also incorporates modern front-end technologies, ensuring an intuitive and responsive user interface. This stack enables enhanced expressiveness and flexibility, allowing even end-users to extend the tool’s functionalities. Using Docker\footnote{\url{https://docker.com/}}, \jjodel{} containerizes its components, ensuring scalable deployment across diverse infrastructures while maintaining consistency and seamless performance.

\begin{figure}
    \centering
\includegraphics[width=0.8\linewidth]{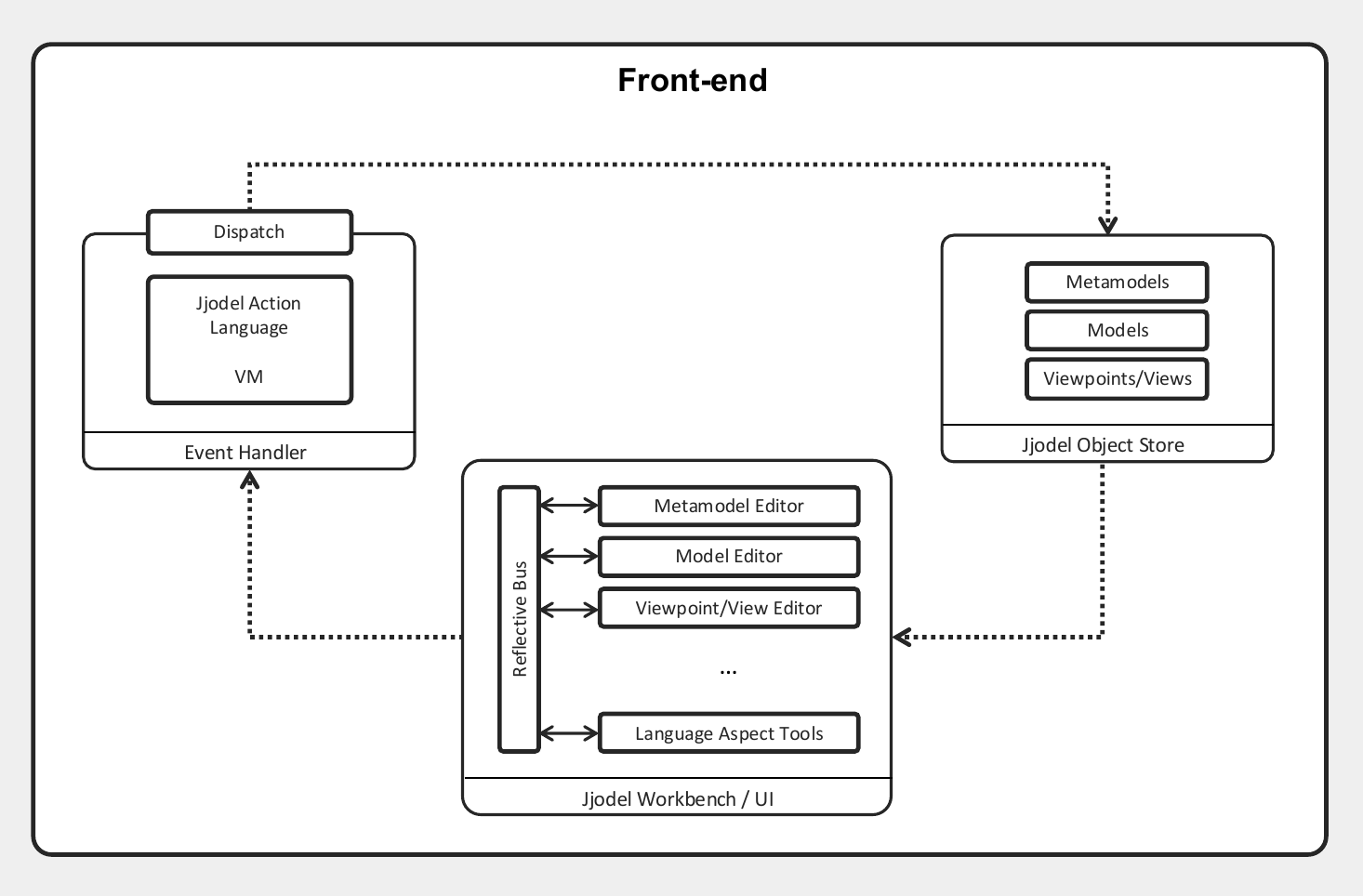}
    \caption{Front-end architecture of \jjodel{}.}
    \label{fig:fe-architecture}
\end{figure}  


%
%
%

\medskip\noindent
The \textit{front-end} serves as the central component for defining and managing MDE projects. Built with TypeScript\footnote{\url{https://www.typescriptlang.org/}} to ensure scalability and maintainability, it leverages React\footnote{\url{https://react.dev/}} to power its dynamic and interactive user interface. It enables users to seamlessly interact with metamodels, models, and concrete syntaxes, creating an intuitive and powerful integrated environment.
The core of the front-end is the \textit{\jjodel{} Object Store}, a centralized repository implemented using Redux\footnote{\url{https://redux.js.org/}}. This layer manages all artifacts while synchronizing application states and ensuring persistent state management. By adopting a unidirectional data flow design pattern (Fig.~\ref{fig:fe-architecture}), the architecture guarantees a predictable, efficient, and robust data handling throughout the front-end.
More details on JjOM APIs for interacting with the \jjodel{} Object Store are provided in Section~\ref{sec:jom-api}, highlighting how developers can query, manipulate, and extend stored artifacts seamlessly.

The \textit{\jjodel{} Workbench/UI}, a core component of the front-end, serves as the primary interface for users. It offers editors for metamodels, models, and viewpoints, enabling users to define structural elements, ensure model conformance, and explore specific perspectives. A key feature of the platform is its built-in governance, made possible by its reflective architecture. This reflectiveness supports live metamodel/model coevolution~\cite{cicchetti2008automating}, allowing the tool to dynamically adapt editors and models in response to metamodel changes, ensuring seamless synchronization. This capability not only facilitates ongoing maintenance but also enables a test-driven development approach for metamodels~\cite{cicchetti2012test}.
In addition, the workbench integrates tools for managing DSL aspects, including model validation, interactive queries, and console functionalities, further improving the modeling experience and improving user productivity.

\medskip
\noindent
The \textit{back-end} manages data persistence within the \jjodel{} repository and enables collaborative modeling in real-time. It is made up of two primary components:
\begin{itemize}
    \item The \textit{\jjodel{} Project Repository}: it handles the storage of project data, including metamodels, models, concrete syntaxes and user settings. It is built on Node.js\footnote{\url{https://nodejs.org/}} for server-side operations, Express.js\footnote{\url{https://expressjs.com/}} for routing and APIs, and MongoDB\footnote{\url{https://www.mongodb.com}} for reliable and scalable data persistence.
	\item The \textit{\jjodel{} Collaborative Service}: designed to facilitate real-time collaboration, this service allows multiple users to interact simultaneously within the platform. It uses Socket.IO\footnote{\url{https://socket.io/}} for real-time communication and data synchronization, with MongoDB ensuring persistent storage of collaborative data. RESTful API endpoints further enhance functionality by managing rooms and coordinating activities during collaborative sessions.
\end{itemize}

Together, these services provide robust back-end support, enabling seamless collaboration and efficient data management, essential for powering the dynamic and interactive features of \jjodel{}.

\subsection{\jjodel{} Object Model (JjOM)}\label{sec:jom-api}
%
The \jjodel{} Object Model (JjOM) defines how modeling artifacts are represented and accessed within the \textit{\jjodel{} Object Store}. It provides a structured way of interaction with model elements through its API, enabling users to query, evaluate, and manipulate models. JjOM also facilitates the definition of concrete syntaxes and provides a dedicated console for testing and debugging, offering a comprehensive framework for model development and refinement.

It organizes modeling artifacts through three interconnected submodels, each serving a specific purpose in managing and presenting modeling artifacts: 
\begin{itemize}

    \item \textit{Data Submodel} (\code{data}): this submodel is closely tied to the foundational elements of modeling, such as classes, attributes, relationships, constraints, and their corresponding instances. These elements form the core building blocks of the \jjodel{} meta-metamodel, as outlined in Table~\ref{tab:core-modeling}, which includes key constructs such as \code{DModel}, \code{DClass}, \code{DAttribute}, \code{DReference}, \code{DObject}, and \code{DValue}. Together, these constructs establish the structural and semantic framework necessary for defining and managing both metamodels and their conforming models.

    \item \textit{Node Submodel}  (\code{node}): responsible for managing layout information, this submodel determines the spatial organization and connectivity of elements within the modeling environment. It also serves as a flexible space for storing additional information used in operations. For example, the validation mechanism uses the state of the node to store and manage validation messages. Because \code{node} is interconnected with the other submodels, it enables the definition of viewpoints and semantic aspects that may depend on the layout information, both to represent and specifying the characteristics of the specific characteristics of the model. For an example, refer to Figure~\ref{fig:expressions};
    
    \item \textit{View Submodel}  (\code{view}): focused on representing the concrete syntax and visual details of model elements, the view submodel synchronizes data and node elements, ensuring consistency between the abstract structure and its visual representation.
\end{itemize}
\begin{table}[h]
    \centering
        \caption{Core Modeling Constructs.}
    \label{tab:core-modeling}
    \begin{tabular}{|l|p{9.5cm}|}
    \hline
        \textbf{Construct} & \textbf{Description} \\ \hline
         \texttt{DModel} & serves as the top-level container holding Packages or Classes. DModel acts as the root of model specification.
  \\ \hline
        \texttt{DPackage} & is a logical grouping or namespace for related classes. \\ \hline
        \texttt{DClass} & represents a class in the metamodel. It can extend from another class and declares, attributes, and references. \\ \hline
        \texttt{DAttribute} & represents a field or characteristic that holds an intrinsic value that can be typed as a primitive type, \ie{} integer, string, boolean, etc., or custom-defined enumeration),\\ \hline
        \texttt{DReference} & specifies a relationship between classes and may include cardinality constraints (e.g., one-to-one, one-to-many). DReferences can be containment or non-containment relationships. \\ \hline
        \texttt{DObject} & is an instance of a DClass. It stores runtime values for each DAttribute and links pointing to other objects.\\ \hline
        \texttt{DValue} & is the concrete value assigned to an attribute or data field. Each DObject’s DAttribute has one or more DValues, that can be a scalar (e.g., string, integer, boolean) or an enumeration literal. \\ \hline
    \end{tabular}
\end{table}

These submodels collectively enable JjOM to provide a comprehensive structure to interact with, visualize, and manage modeling artifacts, ensuring precision, expressiveness, and flexibility. In particular, \code{data}, \code{node}, and \code{view} encapsulate all the information specified by the meta-metamodel and provide programmatic access to these elements through attributes and functions exposed by the JjOM API, summarized in Table~\ref{tab:jjom-api}. This interface allows developers to query, manipulate, and validate modeling artifacts. For example, developers can use the API to retrieve all attributes associated with a specific \code{DClass}, add or remove relationships between \code{DObjects}, or validate constraints defined within a metamodel. 

\newcommand{\tab}{\par\ \ \ }
\newcommand{\nl}{\par}
\begin{table}[p]
\centering\scriptsize
\caption{Attributes and Functions of the JjOM API.}
\label{tab:jjom-api}
\begin{tabular}{|@{\hskip 1mm}p{0.32 \linewidth}|p{0.07\linewidth}|>{\raggedright\arraybackslash}p{0.60\linewidth}|}
\hline
\textbf{Name} & \textbf{Scope} & \textbf{Description} \\ \hline
\multicolumn{3}{|c|}{\textbf{Attributes}} \\ \hline
\code{allInstances: DObject[]} & \code{DModel} & array containing all instances of type \code{DObject} \\ \hline
\code{attributes: DAttribute[]} & \code{DClass} & array of \code{DAttribute}s associated with a \code{DClass} \\ \hline
\code{className: String} & \code{DObject} & the name of the class represented as a \code{String} \\ \hline
\code{extendedBy: DClass[]} & \code{DClass} & array of \code{DClass} that extend this class \\ \hline
\code{extends: DClass[]} & \code{DClass} & array of \code{DClass} objects that this class extends \\ \hline
\code{id: Pointer} & \code{Any} & unique identifier represented as a \code{Pointer} \\ \hline
\code{instanceOf: DClass} & \code{DObject} & reference to the \code{DClass} this object is an instance of \\ \hline
\code{instances: DObject[]} & \code{DClass} & array containing all instances of this \code{DClass} \\ \hline
\code{isAbstract: Boolean} & \code{DClass} & boolean indicating whether the class is abstract \\ \hline
\code{isFinal: Boolean} & \code{DClass} & boolean indicating whether the class is final \\ \hline
\code{isInterface: Boolean} & \code{DClass} & boolean indicating whether the class is an interface \\ \hline
\code{isMetamodel: Boolean} & \code{DModel} & boolean specifying if the model belongs to a metamodel \\ \hline
\code{isPrimitive: Boolean} & \code{DClass} & boolean specifying whether the type is primitive \\ \hline
\code{isRootable: Boolean} & \code{DClass} & boolean specifying whether the class can be a root element \\ \hline
\code{isSingleton: Boolean} & \code{DClass} & boolean indicating whether the class is a singleton \\ \hline
\code{name: String} & Global & the name of the element represented as a \code{String} \\ \hline
\code{objects: DObject[]} & Global & array containing all \code{DObject} instances in the model \\ \hline
\code{operations: DOperation[]} & \code{DClass} & array of operations (\code{DOperation}) defined for the class \\ \hline
\code{packages: DPackage[]} & \code{DModel} & array of \code{DPackage} elements associated with the model \\ \hline
\code{parent: DObject} & \code{DObject} & reference to the parent object, which can be of any type \\ \hline
\code{references: DReference[]} & \code{DClass} & array of references (\code{DReference}) defined for the class \\ \hline
\code{value: DValue} & \code{DValue} & single-valued data for attributes or references with an upper bound of 1 \\ \hline
\code{values: DValue[]} & \code{DValue} & array of multi-valued data for attributes or references \\ \hline
\code{\$<name>: DValue} & \code{DClass}, \code{DModel} & dynamically accesses child elements or instances by exact name \\ \hline
\multicolumn{3}{|c|}{\textbf{Functions}} \\ \hline
\code{addAttribute(}\tab\code{attribute: DAttribute}\nl\code{): DClass} & \code{DClass} & adds an attribute (\code{DAttribute}) to the specified \code{DClass} \\ \hline
\code{addClass(\tab name: String\nl ): DPackage|DModel} & \code{DModel, DPackage} & adds a \code{DClass} with the specified name to a \code{DPackage} or \code{DModel} \\ \hline
\code{addObject(\tab param: JSON,\tab className: String\nl ): DModel} & \code{DModel} & creates an instance of the specified \code{DClass} and initializes it with values from the JSON \code{param} \\ \hline 
\code{addRelationship(\tab type:DRelationType\nl ): DClass} & \code{DClass} & creates a relationship of the specified type between two \code{DClass} entities \\ \hline 
\code{delete (): void} & 
\code{Any}& deletes the current element and cleans up associated references or links to maintain model consistency \\ \hline
\code{executeQuery (\tab query: JSX~Expression\nl ): Any[]} & Global & executes a custom query and returns the matching elements \\ \hline
\code{getAttributes (): DAttribute[]} & \code{DClass} & retrieves the attributes associated with the specified \code{DClass}. \\ \hline
\code{getElementById(id:Pointer}\code{): Any} & \code{DObject} & retrieves an element by its unique identifier \\ \hline
\code{getRelationships():\tab DReference[]} & \code{DClass} & retrieves all relationships connected to the specified \code{DClass} \\ \hline
\code{removeAttribute(\tab attr: DAttribute\nl ): DClass} & \code{DClass} & removes the specified attribute from the given \code{DClass} \\ \hline 
\code{removeRelationship (\tab ref: DRereference\nl ): DClass} & \code{DClass} & deletes the specified relationship from the model \\ \hline 
\code{validateModel(): Error} & \code{DModel} & validates the model against constraints and returns any validation errors \\ \hline
\end{tabular}
\end{table}
%
%
%
 %
%
%
%
In essence, the API offers methods for retrieving and manipulating model elements, their layout configurations, and their syntactic representations. 
The API plays a central role in enabling JSX\footnote{\url{https://legacy.reactjs.org/docs/introducing-jsx.html}} expressions, which are particularly useful for defining navigation paths and constraints within the \jjodel{} environment. By allowing users to define customized constraints and navigational logic, the API facilitates flexible model querying to be used within JSX templates. This versatility not only enhances the usability of the platform, but also allows users to tailor their interactions to specific modeling needs and scenarios. 

\begin{figure}
    \centering
    \includegraphics[width=0.8\linewidth]{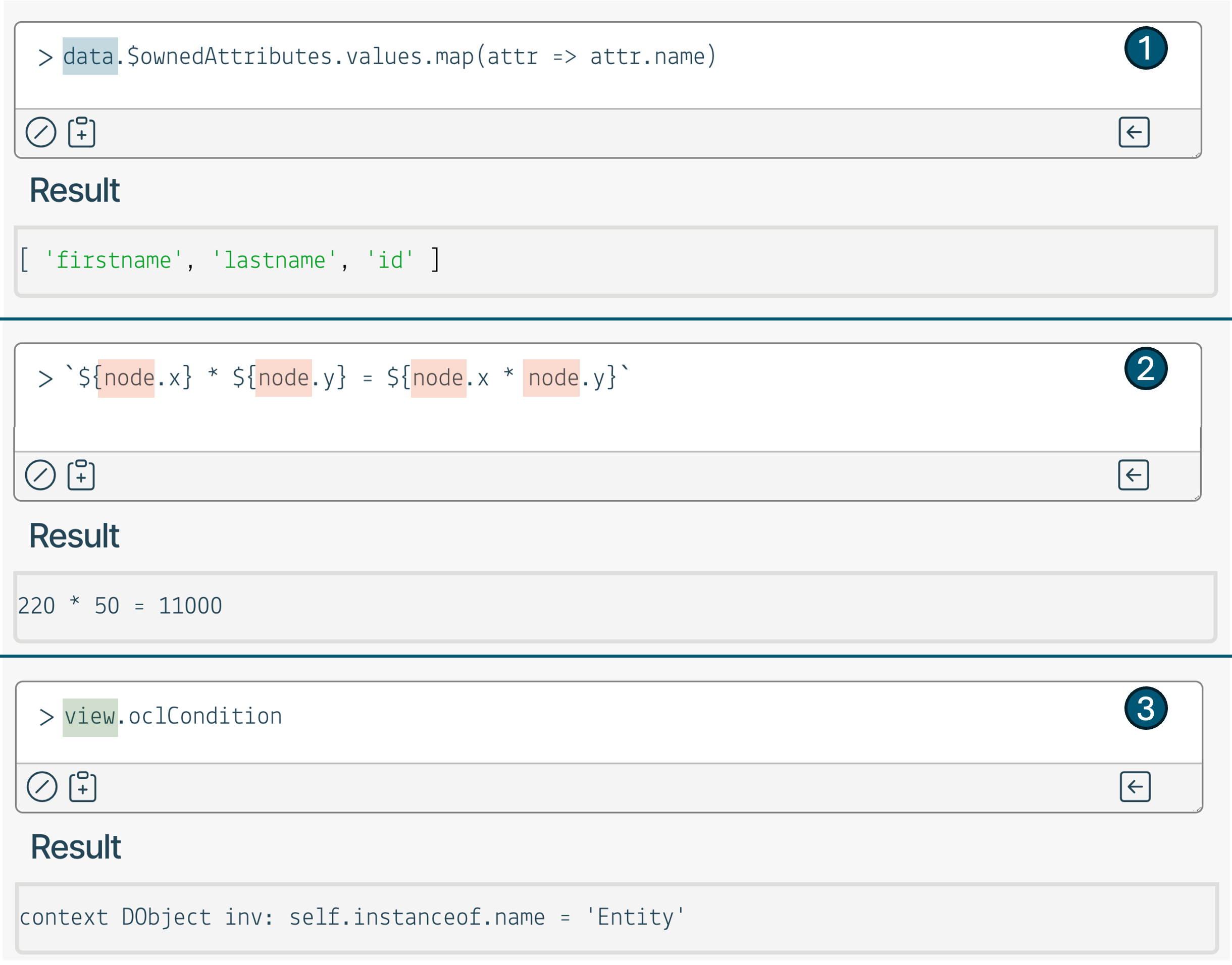}
    \caption{Interacting with JjOM.}
    \label{fig:jjom}
\end{figure}

\medskip
Figure~\ref{fig:jjom} shows three expressions of the JSX exemplar that interact with the submodels \code{data}, \code{node} and \code{view} (highlighted blue, red, and green, respectively), demonstrated within the context of the ERD scenario. The figure illustrates the interactive console of \jjodel{}, which facilitates dynamic exploration and manipulation of these submodels.
%
Each interaction with the console is applied to the selected model element in the editor. In the following, we describe simple statements that interact with each submodel on the \texttt{User} entity:

\begin{itemize}
    \item The expression in console \circled{1} queries \code{data} to retrieve the attributes owned by the \code{User} entity and maps them to their \texttt{name} properties, demonstrating how declarative JSX expressions can extract and manipulate core data elements. In particular, the special character \code{\$} facilitates navigation through the model elements, allowing access to the child elements or instances by their exact name. For example, when an entity instance is selected in the editor, the expression \code{data.\$ownedAttributes} retrieves all the attributes it contains. Essentially, \code{data} functions as the equivalent of \texttt{self} in the OCL language~\cite{Cabot2012}, serving as a contextual reference point for querying and navigating the model elements.
    \item The expression \circled{2} interacts with \code{node}, which manages spatial layout and positioning. The properties \code{node.x} and \code{node.y} represent the coordinates (or other numeric attributes) of a node, corresponding to the visual representation of a model element. The expression multiplies these values, demonstrating the dynamic retrieval and manipulation of layout-related data at run-time.
    \item  The expression \circled{3} references \code{view}, which govern the concrete syntax and visual constraints of the model elements. The \code{oclCondition} enforces that any \code{DObject} in this view must be an instance of \code{Entity}. This shows how OCL-like constraints can be applied to validate or control visual representations on the model front-end. These constraints integrate seamlessly into JavaScript or JSX workflows, enabling UI-level validations and custom conditional rendering.
\end{itemize}
These examples illustrate the expressiveness and versatility of the JjOM API\footnote{A reference guide for the JjOM API is available online (\url{https://www.jjodel.io/jjodel-object-model-jjom/}).} to support advanced modeling scenarios, enabling developers to tightly integrate model data with dynamic and interactive front-end environments.

To further enhance the general usability, a set of predefined facilities has been introduced, facilitating common tasks and reducing the effort required for developers to interact with the JjOM. These facilities provide high-level abstractions and helper functions that simplify complex operations, such as querying model elements, managing layout configurations, and enforcing visual or semantic constraints. Table~\ref{tab:jjom-api} presents a subset of these predefined facilities, showcasing their role in accelerating development workflows and enabling intuitive interactions with \code{data}, \code{node}, and \code{view}. 

In particular, the adopted approach leverages React JSX, a notation widely covered in undergraduate courses, ensuring that many students and professionals are already familiar with its syntax and principles. This familiarity significantly reduces the learning curve, allowing users to quickly adapt and focus on using the features of the framework. Together, these facilities improve efficiency, accessibility, and user experience, helping to make the framework more transparent and as such suitable for a broad audience, from novice developers to seasoned professionals.

\subsection{Syntax Viewpoints}
Syntax viewpoints in \jjodel{} define the concrete syntax of models, specifying how instances of abstract metamodel concepts are visually represented and interacted with. Each viewpoint consists of a collection of \textit{views} that define the mapping between abstract elements and their corresponding concrete representations. This bidirectional mapping ensures not only the visualization of abstract elements, but also, in the case of projectional editing, enables users to directly interact with the concrete representation to modify the underlying abstract model. These views are specified as graphical rules, each tailored to handle a specific class or element type. Each view consists of multiple aspects, enabling precise and intuitive model manipulation by addressing different facets of the visual representation and interaction. In particular,

\begin{itemize}
    \item \textit{ApplyTo}, specifies the predicate used to select the instances to which the view applies; the predicate can be defined using OCL or JavaScript.  
    \item \textit{Template}, defines a JSX template used to specify a component that visually represents each instance that satisfies the predicate defined in the \textit{ApplyTo} configuration. Templates access and manipulate model data using the JjOM API, i.e., declarative and functional JSX expressions, enabling dynamic and flexible visualizations. Additionally, a template can leverage predefined components provided by the \jjodel{} Syntax Definition (JjSD) library, which are designed to facilitate the creation of graphical syntaxes. An excerpt of these JSX components, along with their specific purposes, is listed in Table~\ref{tab:jdl-component}, offering modelers a robust foundation to define intuitive and expressive syntax representations.
    \item \textit{Style}, defines customizable parameters to tailor the visual appearance of graphical components, leveraging CSS/LESS stylings. Language engineers can specify styling variables, such as colors, border properties, and thickness, to ensure consistent and adaptable designs. As shown in Figure~\ref{fig:style-editor}, these parameters are configured through an intuitive interface that provides options to select color palettes, adjust numerical values (e.g. thickness), and define paths for shapes. This flexibility allows modelers to create visually cohesive and customizable syntaxes tailored to their specific requirements.
    \begin{figure}[ht]
    \centering
    \includegraphics[width=0.8\linewidth]{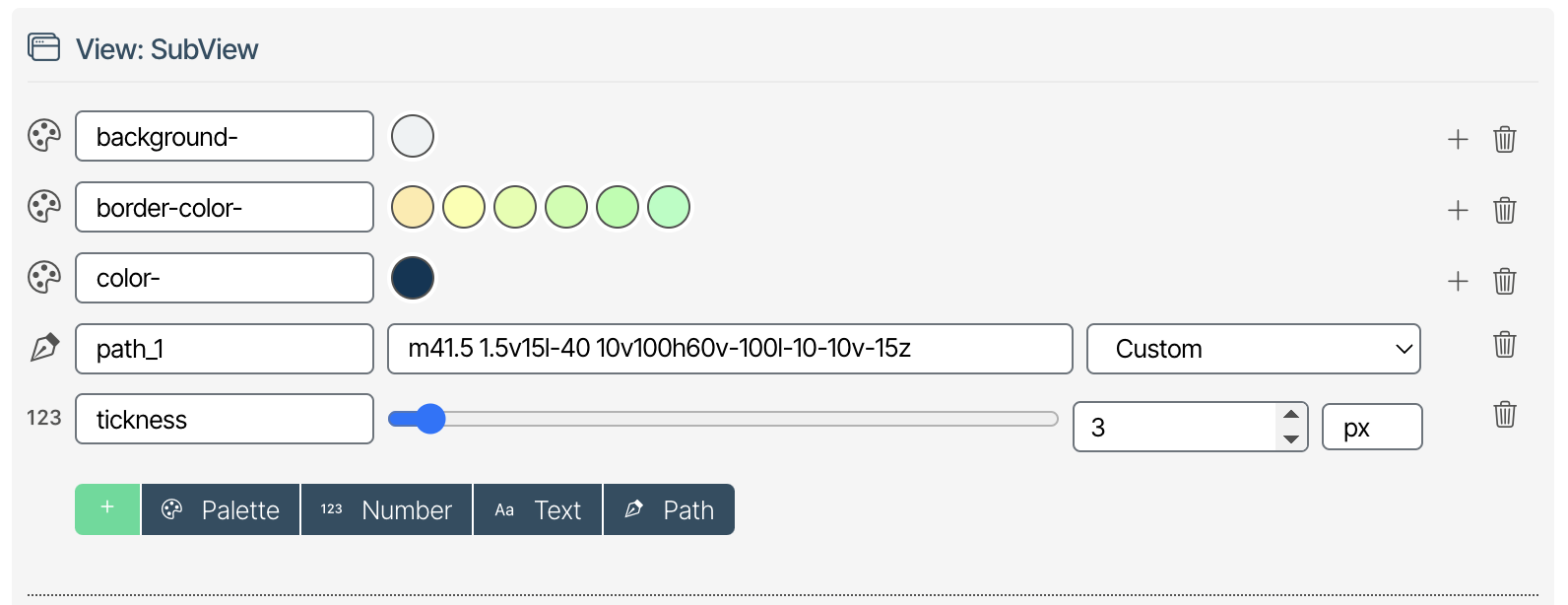} 
    \caption{Example user-defined parameters like color, thickness, and SVG paths for customizing the styling.}
    \label{fig:style-editor}
    \end{figure} 
    \item \textit{Events}, each view allows the modeler to define event handlers as part of Event-Condition-Action (ECA) rules, enabling the capture of user interactions such as dragging, resizing, and other actions. These events enhance interaction, allowing modelers to customize and extend the editor’s behavior while dynamically manipulating the model. Furthermore, events can be used to define the semantics of the modeling notation, linking user actions to meaningful model transformations or updates. A complete list of available event handlers is shown in Table~\ref{tab:trigger}.
    \item \textit{Options}, enables fine-tuning and advanced customizations, allowing modelers to define how elements are structured and aggregated. For example, elements can be configured to appear as a list within containers or as vertices in a graph. These options provide flexibility in organizing and presenting model components, ensuring that the visual representation aligns with the desired semantics and structure of the model.
\end{itemize}

\begin{table}[]
    \footnotesize
    \centering
    \caption{\jjodel{} JSX Reusable Components.}
    \label{tab:jdl-component}
    \begin{tabular}{|l|p{8.5cm}|}
         \hline
\textbf{JjDL Component} & \textbf{Description} \\ \hline
\code{<Control/>} & Adds a control panel to parameterize the syntax rendering. \\ \hline
\code{<DefaultNode/>} & Corresponds to the view definition of an instance specified as an argument. \\ \hline
\code{<Edge/>} & Corresponds to an edge among two nodes. \\ \hline
\code{<Input/>} & Permits the projectional editing of an attribute. \\ \hline
\code{<Selector/>} & Permits the projectional editing among a set of elements. \\ \hline
\code{<Slider/>} & Adds a range slider to define the value of an integer parameter. \\ \hline
\code{<Toggle/>} & Adds a toggle for a boolean parameter. \\ \hline
\code{<View/>} & It is the opening/closing tag for views. \\ \hline
    \end{tabular}
\end{table}

Using the ERD scenario, the listings~\ref{lst:entity-template}, \ref{lst:attribute-template}, and \ref{lst:relation-template} illustrate how the templates for entities, attributes, and relations are specified within the example ERD context. Each component plays a distinct role in rendering and editing model elements in a 'projectional' style, ensuring a dynamic and customizable user experience.
The component \code{View} serves as the root or container for a particular visualization, encapsulating the layout of all templates, such as entity, attribute, and relation templates. Within this structure, the \code{} component, a key element of the \jjodel{} Syntax Definition (JjSD) library, plays a pivotal role in rendering sub-elements (often instances). It is a generic and polymorphic component that ensures that subelements are consistently represented, even if no custom view is explicitly defined for them.
In the entity template, attributes are iterated using the \texttt{ ownedAttributes} property of the entity, and each attribute is rendered using \code{}. This component inductively identifies the appropriate view to apply by performing a run-time lookup based on the value of the instance that satisfies the view predicates. Since the correct view cannot always be determined statically, this dynamic mechanism allows \code{} to flexibly adapt to different scenarios and contexts. It acts as a bridge, linking to a corresponding model template or providing a minimal default rendering when no custom view is specified.
For the ERD scenario, the \code{} component for attributes is explicitly bonded to the \texttt{Attribute} template. This ensures a seamless and consistent rendering of attributes within the entity visualization while leveraging the generic and polymorphic capabilities of the component to dynamically identify and apply the appropriate view.

All three templates utilize the \code{Input} component to enable projectional (in place) editing of a single attribute value, such as a name or a type. In the \code{Entity} template, \code{Input} is bound to \code{data.\$name}, representing the entity’s name. The property \code{field={‘value’}} specifies that it edits a single textual value. The component \code{Input} also supports additional properties; for example, the property \code{autosize} allows the text box to dynamically resize to fit its content, improving usability and visual clarity.
The \texttt{Attribute} template extends this functionality with the inclusion of the \texttt{Selector} component, which provides a dropdown list to select an enumeration literal, such as the \texttt{type} of an attribute. This enables intuitive interaction and simplifies the selection process for users.
The \texttt{Edge} component plays a critical role in representing connections between nodes in a graphical notation, such as relationships between entities in a diagram. In the ERD example, the \texttt{Relation} template defines two edge elements to visually represent how the \texttt{Relation} node connects to its associated left and right entities. Specifically, the first edge of the \texttt{Relation} node acts as the 'start', while the reference entity node serves as the 'end'. This means that the edge is drawn from the visual element of the relation to the entity it refers to. By specifying which nodes to connect, the modeler can draw edges between multiple nodes in the model (e.g., from a \texttt{User} entity to a \texttt{Role} entity), enabling clear and expressive visualizations of relationships.
\begin{lstlisting}[language=JSX-template,caption=ERD Entity template.,label=lst:entity-template]
<View className={'entity'}>
    <div className={'entity-header'}>
        <div className={'input-container mx-2'}>
            <Input
                data={data.$name}
                field={'value'}
                hidden={true}
                autosize={true}
                style={{backgroundColor: 'transparent', border: 'none'}}
            />
        </div>
    </div>
    <div className={'entity-body'}>
        {data.$ownedAttributes &&
            data.$ownedAttributes.values &&
            data.$ownedAttributes.values.map((attribute) => (
                <DefaultNode data={attribute} key={attribute.id} />
            ))}
    </div>
    {decorators}
</View>
\end{lstlisting}
\begin{lstlisting}[language=JSX-template,caption=ERD Attribute template.,label=lst:attribute-template]
<View className="attribute">
    <div className="attribute-header">
        <div className={'input-container mx-2'}>
            <Input
                data={data}
                field={'name'}
                hidden={true}
                autosize={true}
                style={{ backgroundColor: 'transparent', border: 'none', textAlign: 'center' }} />
            {data.$isPK.value  && '(PK)'}
            <Selector data={data} field={'type'}/>
        </div>
    </div>
    {decorators}
</View>
\end{lstlisting}
\begin{lstlisting}[language=JSX-template,caption=ERD Relation template.,label=lst:relation-template]
<View className="relation">
    <div className="relation-inner">
        <div className="relation-header">
            <div className={'input-container'}>
                <Input
                    data={data.$name}
                    field={'value'}
                    hidden={true}
                    autosize={true}
                    style={{ backgroundColor: 'transparent', border: 'none', textAlign: 'center' }}
                />
            </div>
        </div>
    </div>
    <Edge
        view={'EdgeAssociation'} 
        key={`${data.id}-${data.$left.value.id}`}
        start={node} 
        end={data.$left.value.node} 
    />
    <Edge
        view={'EdgeAssociation'} 
        key={`${data.id}-${data.$right.value.id}`}
        start={node} 
        end={data.$right.value.node} 
    />
    {decorators}
</View>
\end{lstlisting}


\subsection{Validation viewpoint} 
\begin{figure}
    \centering
    \includegraphics[width=0.9\linewidth]{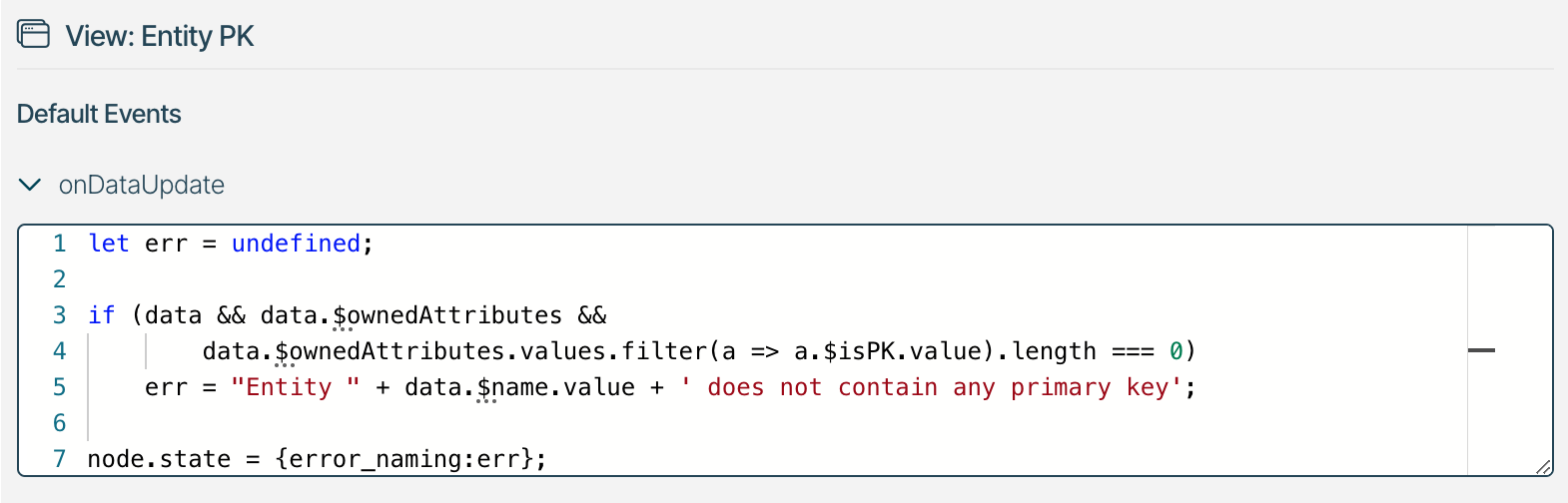}
    \caption{Validation rule.}
    \label{fig:validation}
\end{figure}

The validation viewpoint enforces the rules and constraints defined in the metamodel, ensuring the general consistency of the model and structural correctness. By enabling proactive error detection and providing immediate feedback during the modeling process, it helps reduce errors and reinforces best practices.
As a specialized viewpoint, the validation viewpoint focuses on identifying invalid elements within a modeling artifact and alerting users when these elements violate established constraints. This direct feedback mechanism simplifies the identification and resolution of issues, ensuring that the model remains robust and accurate. In \jjodel{}, validation rules are defined using views to ensure model consistency and identify errors. The \texttt{onDataUpdate} configuration is crucial to evaluating the validity of model instances, while the \texttt{state} property of a node object serves as a repository for storing validation errors associated with the model. These errors are seamlessly tied to the relevant model elements, enabling precise and context-sensitive feedback to guide modelers in solving issues.
In the ERD scenario, Figure~\ref{fig:erd}(b) illustrates a validation marker triggered when an entity lacks a primary key attribute. To implement this validation rule, the \texttt{onDataUpdate} handler is used to evaluate the model and identify any violations. Validation errors are stored in the \texttt{state} property of the corresponding node, ensuring that errors are persistently tied to the affected elements.
Figure~\ref{fig:validation} demonstrates how to enforce the constraint of requiring at least one primary key for each entity. The implementation initializes an \texttt{err} variable to \texttt{undefined}. Using the JjOM framework, model elements are navigated to check if at least one of the \texttt{ownedAttributes} is marked as a primary key. If no primary key is found, the \texttt{err} variable is updated with an error message. Finally, the error message is stored in the \texttt{state} property of the corresponding node. If a node contains an error message, a validation marker (e.g., an alert dialog) is displayed near the affected entity, as shown in Figure~\ref{fig:erd}(b).

This approach ensures that modelers receive immediate visual feedback, enabling them to address validation errors efficiently and maintain the integrity of the modeling artifact.

\begin{table}[b!]
\centering
\caption{Rule triggers.}
\label{tab:trigger}
\begin{tabular}{|p{2cm}|p{6cm}|}\hline
\textbf{Trigger} & \textbf{Description}\\\hline
{onDataUpdate} & Triggered when model data changes.\\\hline
{onDragStart} & Triggered when a drag action starts.\\\hline
{whileDragging} & Triggered during a drag action.\\\hline
{onDragEnd} & Triggered when a drag action ends.\\\hline
{onResizeStart} & Triggered when a resize action starts.\\\hline
{whileResizing} & Triggered during a resize action.\\\hline
{onResizeEnd} & Triggered when a resize action ends.\\\hline
\end{tabular}
\end{table}

\section{Flexibility and Flexible Modeling}\label{sec:advanced}









A core objective of \jjodel{} is to offer flexibility, enabling end users and language engineers to navigate the complexity and richness of modern multifaceted domains. Flexibility is not just a feature, but a foundational principle embedded within its design and technology stack. This approach allows language designers to tailor the modeling environment to their specific needs without encountering the challenges typically associated with custom coding or inflexible extension mechanisms.

To show the flexibility of \jjodel{} in action, we present two illustrative examples: \textit{grid snapping} and \textit{semantic zooming}~\cite{frisch2008towards}. These examples highlight how \jjodel{}'s design facilitates effortless customization of modeling functionalities, demonstrating its adaptability to diverse user requirements and complex scenarios.

\subsection{Grid-snapping}\label{grid-snapping}

Grid snapping is an essential feature in modeling environments, enhancing precision, usability, and efficiency. Although its implementation can be challenging, requiring accurate layout information and dynamic adaptability, these challenges are outweighed by the benefits it provides. By automating alignment tasks, grid snapping allows users to concentrate on the creative and semantic dimensions of their models, ensuring both functional accuracy and aesthetic refinement.

\jjodel{} offers a flexible framework for extending modeling editors by integrating two powerful methodologies: (i) JSX templating and CSS for advanced visual customization, and (ii) Event-Condition-Action (ECA) rules~\cite{widom1995active} for introducing dynamic behavior. As an example, implementing a dot-based grid—a visually minimal yet effective tool for aligning and organizing elements within the modeling editor—can be accomplished through the following steps: 

\begin{enumerate}
\item Defining a CSS class to render the grid as a pattern of dots on the editor canvas.  
\item Adding a toggle command to dynamically enable or disable the grid.  
\item Specifying an ECA rule to snap elements to the nearest grid vertex during movement.  
\end{enumerate}

Each of these steps is elaborated in the following paragraphs.

\paragraph{Step 1: Defining the Grid Style.}

The dot-based grid style is implemented through a simple CSS class definition, as shown below:

\begin{lstlisting}[language=css]
 .grid {
    background-image: radial-gradient(silver 1px, transparent 0);
    background-size: 15px 15px; /* 15px x 15px grid */
 }
\end{lstlisting}

This class renders the grid as a pattern of evenly spaced dots. In the next step, we show how to dynamically apply or remove the \code{.grid} class in the model view.

\paragraph{Step 2: Adding a Toggle Command for Grid Control.}

\begin{figure}
\centering
\includegraphics[width=0.75\linewidth]{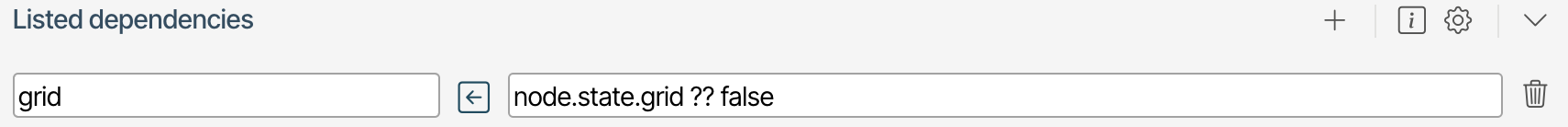}
\caption{The user-defined \code{grid} parameter.}
\label{fig:grid-parameter}
\end{figure}

\jjodel{} enables the extension of the editor functionalities using the \code{<Control/>} component within the model view\footnote{The model view is included in the default viewpoint but can be cloned and customized as needed.}. In this example, a \code{<Toggle/>} component is added to dynamically enable or disable the grid. The toggle operates on a Boolean parameter, \code{grid}, which is defined on the template page, as illustrated in Figure~\ref{fig:grid-parameter}.
The parameter \code{grid} serves as a shortcut for the expression:\\
\begin{lstlisting}[language=JSX-template]
node.state.grid ?? false
\end{lstlisting}
which initializes the parameter in the \code{node} submodel to \code{false}. The updated template for the model view is shown below:
\begin{lstlisting}[language=JSX-template]
 <View className={model ${grid && 'grid'}}>
    ...
    <Control title={'Workbench'} payoff={'Controls'}>
        <Toggle name={'grid'} title={'Grid'} node={node} />
    </Control>
 </View>
\end{lstlisting}
Here, the conditional expression \code{\$\{grid \&\& 'grid'\}} dynamically applies the \code{.grid} class whenever the value of the parameter \code{grid} is \code{true}, allowing the feature.

\begin{figure}[h]
\centering
\includegraphics[width=0.95\linewidth]{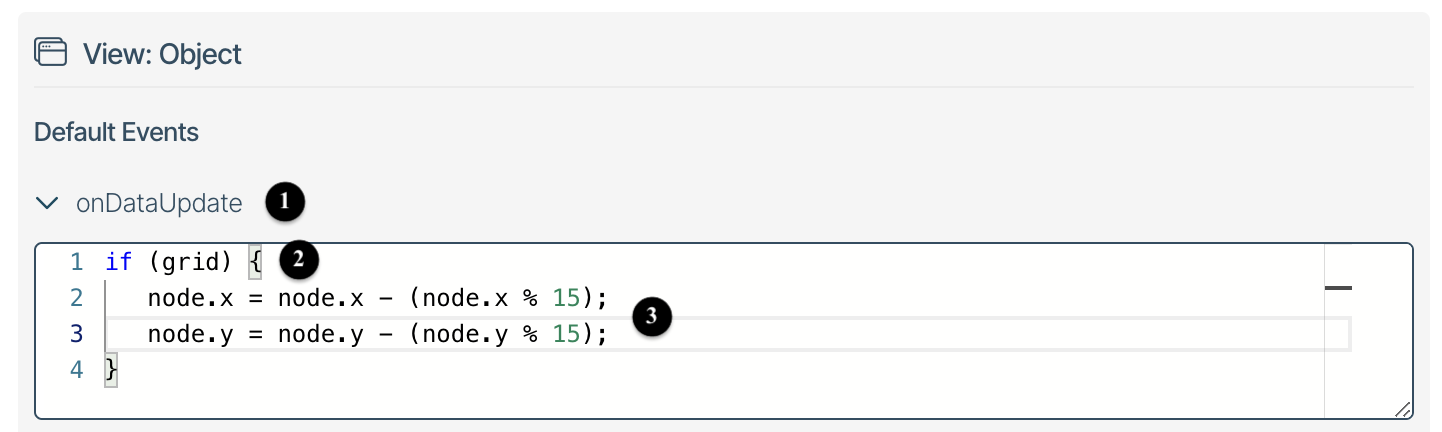}
\caption{ECA rule for the \textit{snap-to-grid} feature.}
\label{fig:eca}
\end{figure}

\paragraph{Step 3: Enabling Snap-to-Grid Behavior.}
To implement the snap-to-grid functionality, an ECA rule is defined, as shown in Figure~\ref{fig:eca}. This rule takes advantage of the \code{grid} parameter to ensure that the elements align with the nearest grid vertex when moved.
The event \code{onDataUpdate} \circled{1} is triggered when an element is moved. Condition \circled{2} verifies the state of the parameter \code{grid}, and if enabled, the action \circled{3} adjusts the coordinates \code{x} and \code{y} of the model element to align with the nearest grid vertex.  

\begin{figure}[h!]
\centering
\includegraphics[width=1\linewidth]{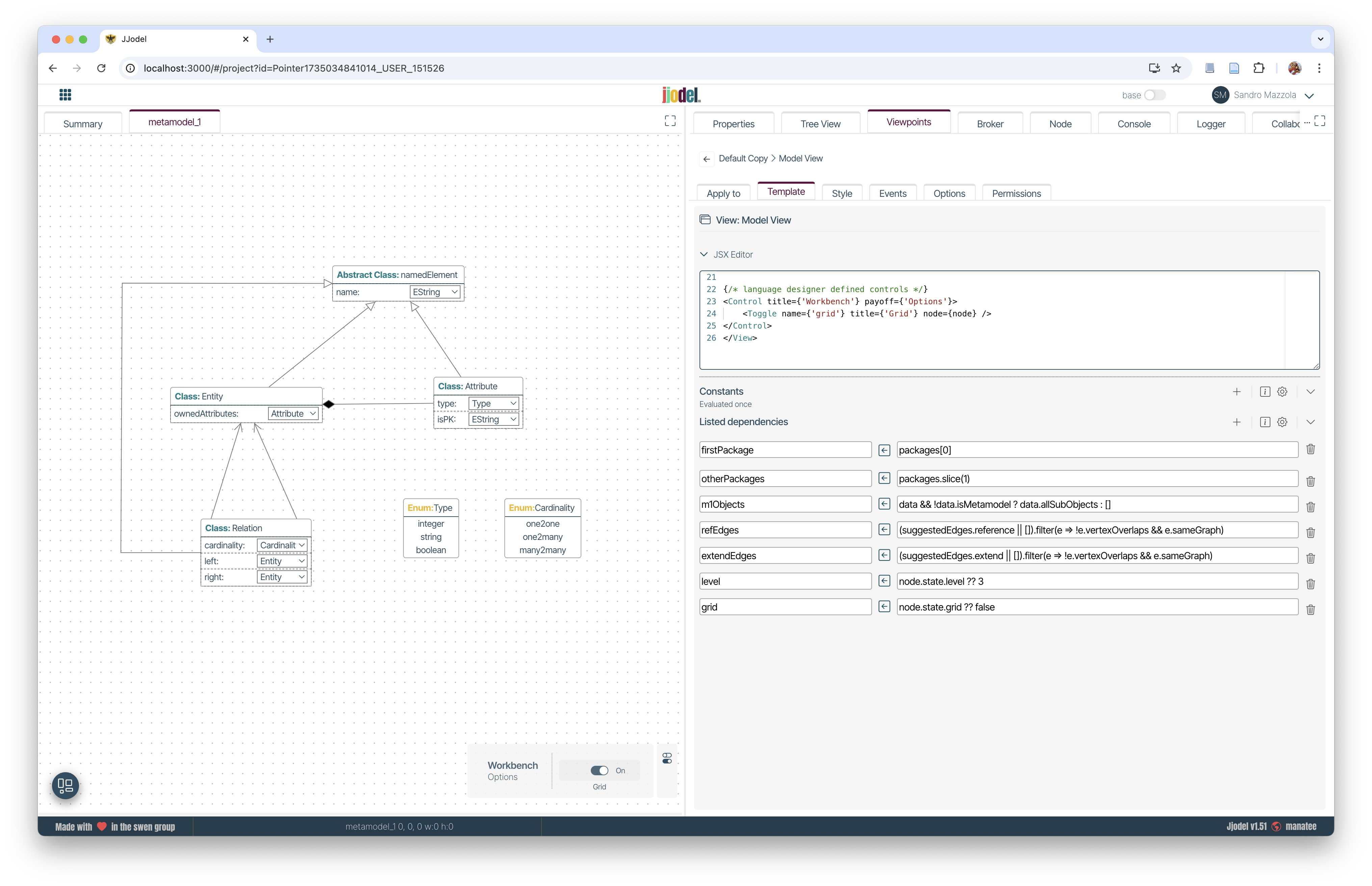}
\caption{Grid-enabled editor.}
\label{fig:editor-grid}
\end{figure}

Figure~\ref{fig:editor-grid} provides a screenshot of the editor with the grid enabled and \code{<Toggle/>} on.

\subsection{Semantic Zooming}
Semantic zooming~\cite{frisch2008towards} is an advanced interaction paradigm that dynamically adjusts the level of detail presented in a model based on the zoom level. Unlike traditional zooming, which merely magnifies or shrinks visual elements, semantic zooming alters the content itself to better align with the user’s context and focus. In modeling, this approach offers several benefits, including enhanced usability, reduced cognitive load, and more efficient navigation by presenting only the most relevant details at each zoom level~\cite{pirolli2001focus}. For example, at higher zoom levels, intricate details of specific model elements become visible, while at lower zoom levels, complexity is abstracted to provide an overview of the entire model.

In \jjodel{}, semantic zooming is achieved by linking different visual representations to predefined zoom thresholds. These representations are dynamically updated as users adjust the zoom level, ensuring that the displayed content remains contextually relevant and appropriately detailed. Although semantic zooming might seem more conceptually complex than features such as grid snapping, it is implemented seamlessly in \jjodel{} thanks to its flexible templating framework. This framework allows for the integration of controls, such as parameters similar to \code{grid} discussed above, that dynamically influence visualization without affecting the underlying models.

Semantic zooming in the default syntax is implemented through the following steps:  

\begin{enumerate}  
\item Define an integer parameter, such as \code{level}, to represent the current zoom level.  
\item Add a slider control to the user interface that allows users to interactively adjust the zoom level.  
\item Divide the template into sections corresponding to different zoom levels, ensuring that the level of detail displayed dynamically adjusts based on the parameter's value.  
\end{enumerate}  
Each of these steps is elaborated in the following paragraphs.

\begin{figure}  
    \centering  
\includegraphics[width=0.75\linewidth]{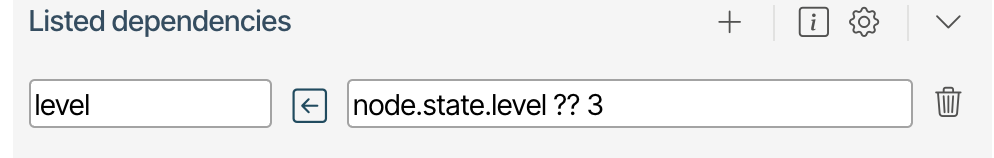}  
    \caption{User-defined \code{level} parameter.}  
    \label{fig:level}  
\end{figure}

\paragraph{Step 1: Defining a new integer parameter.} This step mirrors the process of defining the parameter \code{grid} for the snap-to-grid feature discussed earlier. Here, a user-defined parameter serves as a dynamic control to adjust the visualization. For example, the \code{level} parameter can be introduced in the template as part of the dependency settings. It is initialized using an expression such as:  

\begin{lstlisting}[language=JSX-template]  
node.state.level ?? 3
\end{lstlisting}  
depicted also in Figure~\ref{fig:level}. This ensures that the parameter is stored in the \code{node} submodel and defaults to 3 (or any specified starting zoom level) if the user does not explicitly set it.  

The parameter \code{level} acts as a bridge between user input and the dynamic behavior of the semantic zooming feature. By making the zoom level accessible and modifiable, it allows the modeling editor to adjust the level of detail displayed dynamically based on the current zoom value. This foundational step ensures that the parameter is seamlessly integrated into the template and other relevant parts of the modeling environment, setting the stage for subsequent steps in the implementation of semantic zooming.  

\paragraph{Step 2: Adding a Slider Control for Zoom Level Adjustment.}

\paragraph{Adding a Slider Control for Zoom Level Adjustment.} This step parallels the use of the \code{<Toggle/>} component for the snap-to-grid feature but utilizes a \code{<Slider/>} component to modify the \code{level} parameter.  

The component \code{<Slider/>} is added to the control panel in the template, providing an intuitive mechanism for users to adjust the zoom level. Dynamically updates the \code{level} parameter stored in the \code{node} submodel, ensuring real-time responsiveness. This parameter acts as a bridge between user input and the dynamic behavior of the template.  

For instance, the updated template might look as follows:  

\begin{lstlisting}[language=JSX-template]  
 <View className={`model`}>  
   ...  
   <Control title={'Workbench'} payoff={'Zoom Controls'}>  
      <Slider name={'level'} title={'Zoom level'} node={node} min={0} max={3} />  
   </Control>  
 </View>    
\end{lstlisting}  

In this example:  
\begin{itemize}
    \item In line 4, the \code{name} property links the slider to the \code{level} parameter, ensuring that any adjustments made via the slider directly update the parameter. 
    \item In line 5, additional properties such as \code{min} and \code{max} define the range of zoom levels (e.g., 0 to 3) and the granularity of adjustments.  
\end{itemize}

Similar to the grid-snapping feature, this slider control integrates seamlessly with the \jjodel{} framework, enabling users to dynamically modify visualizations without directly interacting with the model data. The \code{level} parameter remains synchronized with the slider's position, facilitating context-aware rendering based on the zoom level, which will be detailed in the next step.  

\paragraph{Step 3: Slicing the Template for Semantic Zooming.} This step specifies how content dynamically adapts as the zoom level changes, ensuring that the displayed information is contextually relevant and aligned with the user’s focus. The template is divided into sections, each corresponding to a specific zoom level. For example, at lower zoom levels (e.g. \code{level = 0}), the template displays high-level abstractions of the model, while at higher zoom levels (e.g., \code{level = 3}), more detailed information is revealed.  

The following is a simplified example of a template associated with a generic metaclass \code{<metaclass-name>}:  

\begin{lstlisting}[language=JSX-template]  
 <View className={'metaclass-name'}>  
    {level === 0 && (  
        <div className={'overview'}>  
            {/* Render high-level abstractions */}  
            Overview of the model...  
        </div>  
    )}  
    {level === 1 && (  
        <div className={'mid-detail'}>  
            {/* Render mid-level details */}  
            Model with basic elements...  
        </div>  
    )}  
    {level >= 2 && (  
        <div className="full-detail">  
            {/* Render detailed elements */}  
            Complete model details...  
        </div>  
    )}  
 </View>  
\end{lstlisting}  

In this example, lines 2, 8, and 14 use conditional rendering to determine which content is displayed based on the value of the \code{level} parameter. This approach ensures that the visualization dynamically adjusts to the user’s zoom level, providing the appropriate level of detail for their current focus.  

The described approach to slicing the template ensures that semantic zooming is both straightforward to implement and highly effective, creating a dynamic and user-friendly modeling environment. Using the \code{level} parameter and the \jjodel{} templating system, semantic zooming is achieved without the need for custom coding or complex extension mechanisms.  

It is worth noting that this is just one method of implementing semantic zooming, using a single template with multiple sections corresponding to different zoom levels. An alternative approach could involve using a separate view for each \code{level} value, which offers additional flexibility depending on the use case.

\section{\jjodel{} in Practice}
\label{sec:practice}
\jjodel{} is a comprehensive metamodeling framework tailored for language designers to define, extend, and refine DSLs and their associated modeling notations. It offers a structured and systematic approach to capture key structural elements of a language through formalized viewpoints. These viewpoints address critical aspects of language design, including abstract syntax, concrete syntax (textual, visual, or hybrid), validation rules, text generation, and semantics such as editor behavior. Using these capabilities, \jjodel{} facilitates the creation of a precise, flexible, and robust domain-specific workbench, ensuring clarity, consistency, and adaptability throughout the design process.

\begin{figure}[ht]
    \centering
    \includegraphics[width=0.55\linewidth]{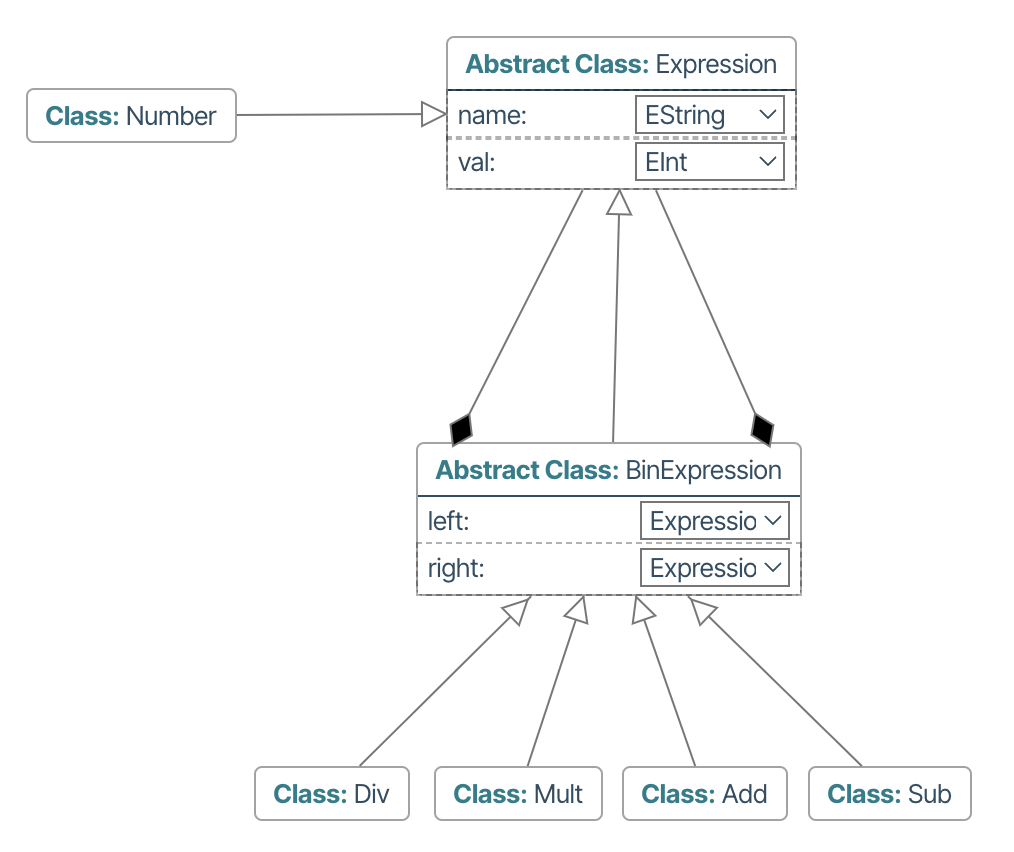}
    \caption{The metamodel of a simple Expression Language.}
    \label{fig:expr-metamodel}
\end{figure}

To illustrate the practical application of \jjodel{}, consider the expression language, a concise example that is often used to introduce students to the foundational concepts of MDE. The abstract syntax of this language is formalized in the metamodel shown in Figure~\ref{fig:expr-metamodel}, which provides a structured framework to represent algebraic expressions. The metamodel consists of abstract classes such as \code{Expression} and \code{BinExpression}, along with their relationships (\code{left} and \code{right}). These components enable the representation of atomic values (\code{Number}) and composite operations (\code{Add}, \code{Sub}, \code{Mult}, \code{Div}). Each expression includes a \code{val} property that stores its computed value, whether representing a literal number or the result of a composite operation. This structure encapsulates the semantics of mathematical expressions while supporting positional representation through concrete syntax definitions.

For instance, consider the following algebraic expressions:
\begin{eqnarray}
    1000 - ((212 + 2) + 102) = 684\\
    ((212 + 2) + 102) - 1000 = -684
\end{eqnarray}

\begin{figure}[ht]
    \centering
    \subfigure[Expression (1)]{
        \includegraphics[width=.32\textwidth]{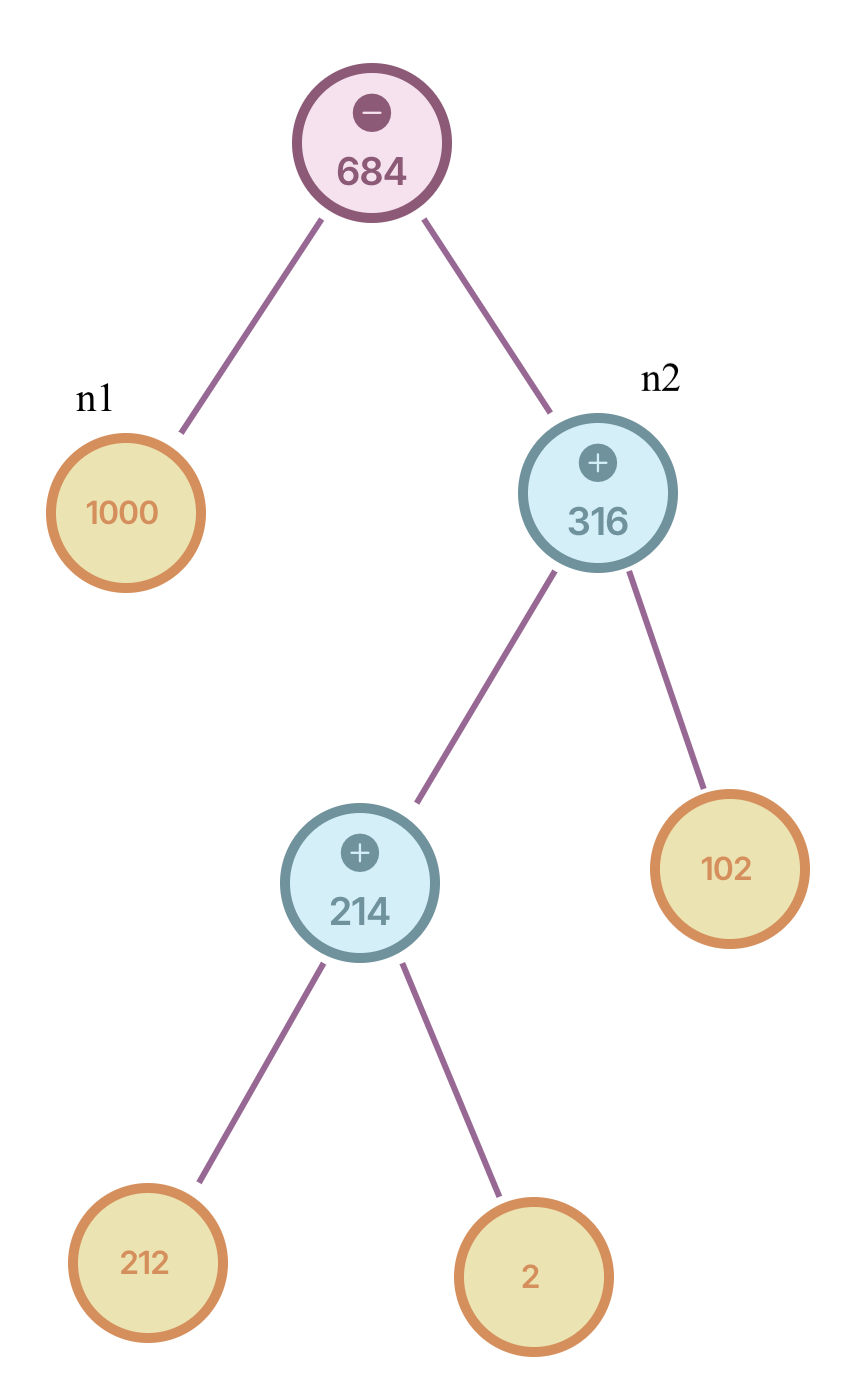} 
    }
    \quad
    \subfigure[Expression (2)]{
        \includegraphics[width=.41\textwidth]{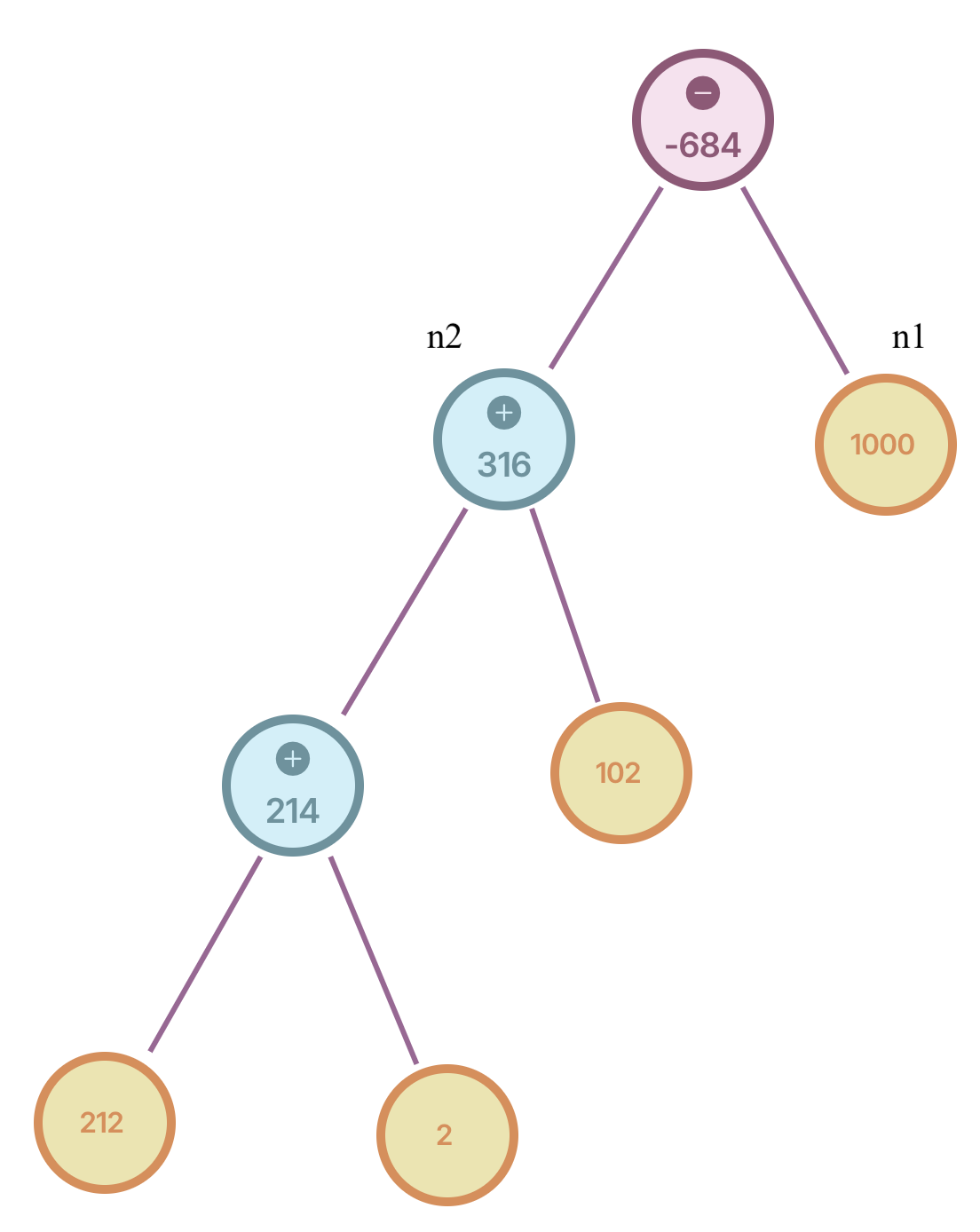} 
    }
    \caption{Expressions illustrating the influence of positional semantics.}
    \label{fig:expressions}
\end{figure}

\begin{figure}[ht]
    \centering
    \includegraphics[width=0.85\linewidth]{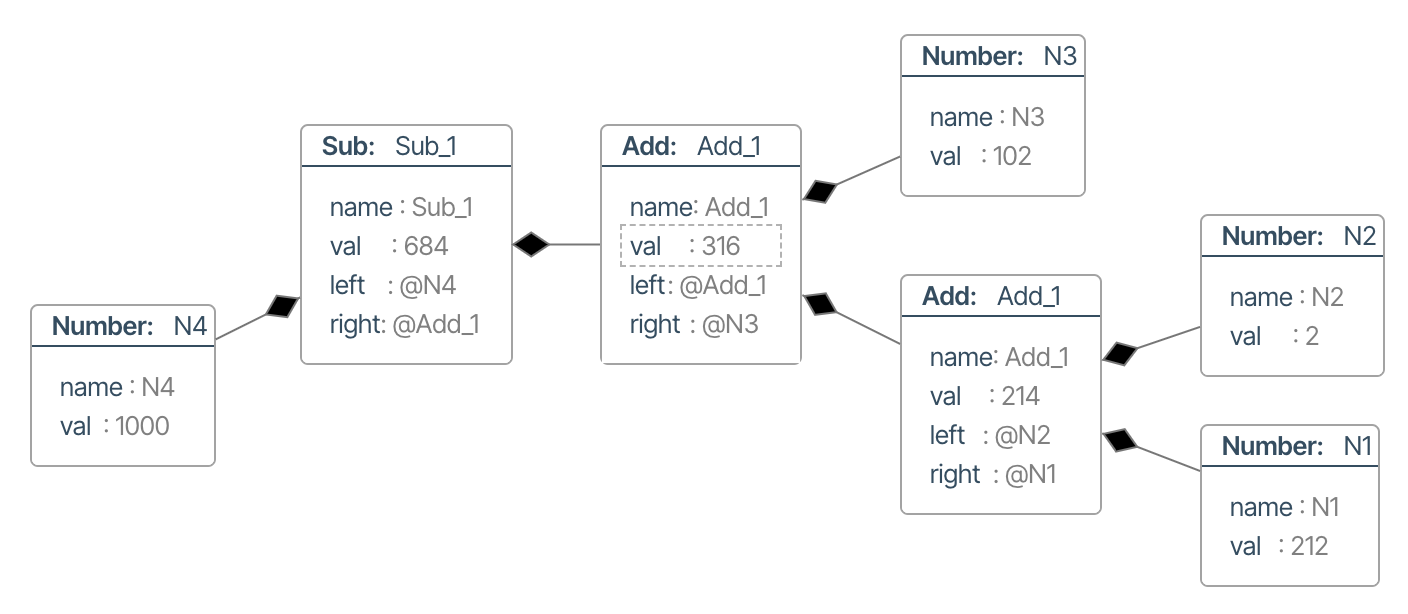}
    \caption{Abstract syntax of expressions (1) and (2).}
    \label{fig:expr-abstract}
\end{figure}

These expressions involve nested binary operations in which subtraction serves as the root operation. Their tree-based representations are shown in Fig.~\ref{fig:expressions}. Although these expressions share the same abstract syntax (Fig.~\ref{fig:expr-abstract}), their computed values differ due to the positional nature of algebraic notation. Subtraction, being non-commutative, depends on operand order (e.g., $(10 - 5) \neq (5 - 10)$). To address this, the concrete syntax \jjodel{}, defined in visual viewpoints, encodes layout information to represent positional semantics.

\jjodel{} leverages its dynamic viewpoints to manage these semantics effectively. For example, the \code{Number} view renders numbers as colored nodes, assigning values to their \code{val} property using JSX templates and CSS rules. Composite operations such as addition (\code{AddView}) and subtraction (\code{SubView}) are governed by the ECA rules~\cite{widom1995active}.

\begin{figure}[ht]
    \centering
    \includegraphics[width=0.95\linewidth]{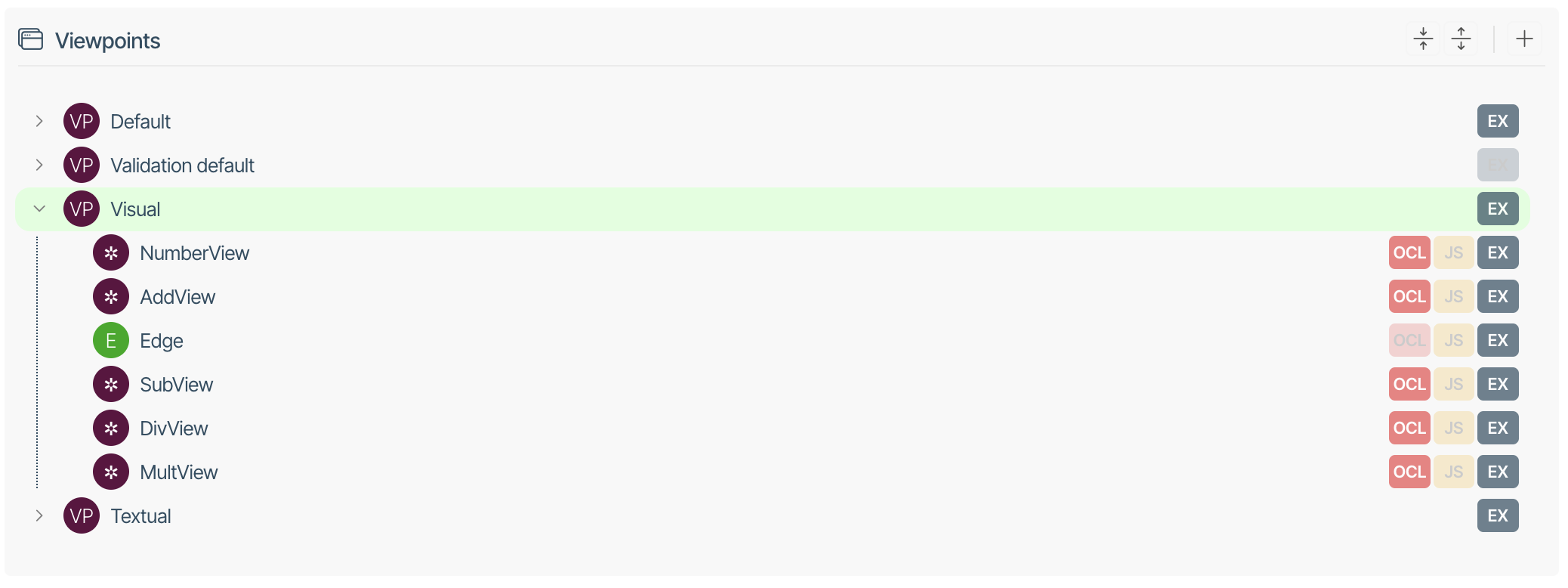}
    \caption{Viewpoints and their corresponding views.}
    \label{fig:vps}
\end{figure}

ECA rules consist of three components:
\begin{itemize}
    \item Event: \code{onDataUpdate}, triggered whenever the data associated with a model element change.
    \item Condition: Defining the OCL predicates and determining the elements to which a view applies.
    \item Action: Specifies how values are computed or updated using features defined in the metamodel (e.g., \code{\$val}, \code{\$left}, \code{\$right}). For instance, addition is defined as follows:
\begin{center}
    \includegraphics[width=1\linewidth]{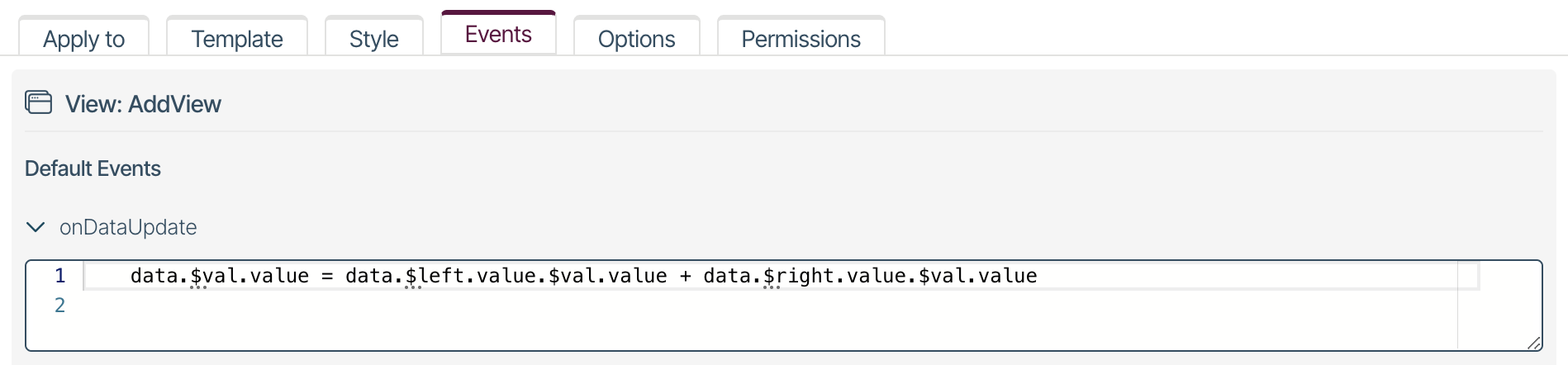}
\end{center}
\end{itemize}

For subtraction, the implementation of \code{SubView} ensures positional semantics by enforcing relative operand positioning:
\begin{lstlisting}[language=JSX-template]
if (data.$left.value.node.x < data.$right.value.node.x) {...}
\end{lstlisting}

The dynamic updates in \jjodel{} propagate seamlessly. For example:
\begin{enumerate}
    \item At $t_0$, the user modifies $e_0$ from $212$ to $112$.
    \item This triggers an \code{onDataUpdate} event, updating $e_1$ at $t_1 > t_0$.
    \item The update cascades to $e_3$, recalculating the value at $t_2 > t_1$.
    \item Finally, $e_4$ is re-evaluated, producing the correct result at $t_3 > t_2$.
\end{enumerate}

\begin{figure}[ht]
    \centering
    \includegraphics[width=.9\linewidth]{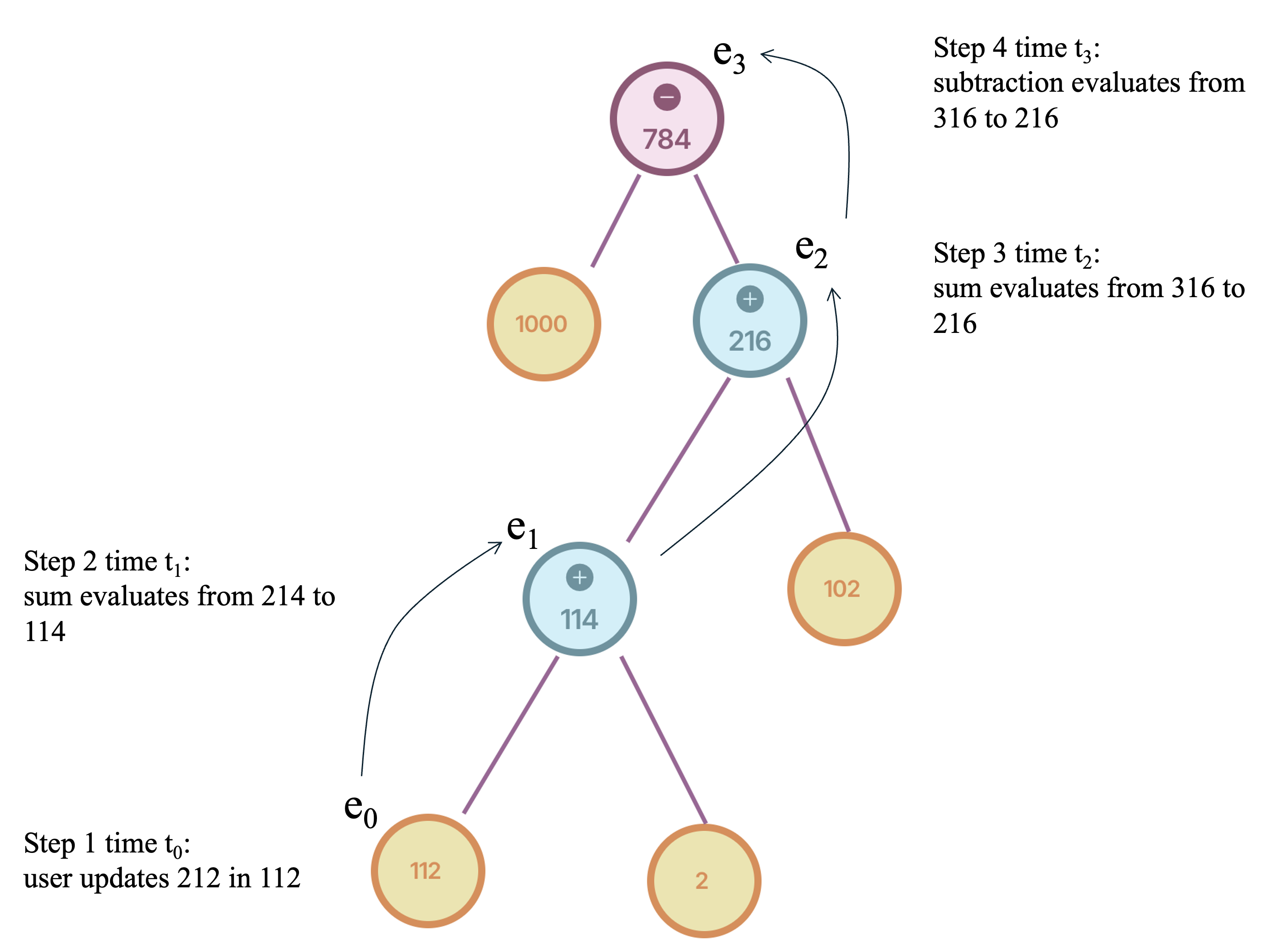}
    \caption{Dynamic update propagation workflow.}
    \label{fig:updates}
\end{figure}

By automating dependency resolution, \jjodel{} ensures consistency and eliminates redundant computations, allowing users to focus on high-level design tasks. This example highlights the capabilities of \jjodel{} in streamlining modeling workflows and providing a powerful and intuitive framework to tackle complex MDE challenges.


\section{Reflections and Lessons Learned}
\label{sec:lessons}
The design and implementation of \jjodel{} provided invaluable insights into the challenges and opportunities of creating a modern and user-friendly modeling platform. These lessons, which span architectural decisions, usability considerations, and technology integration, offer a foundation for advancing the field of MDE tools while reflecting on the successes and setbacks encountered during platform development.

\subsection{The Impact of Technology Choices}

The selection of foundational technologies such as Node.js and React played a key role in shaping the architecture, scalability, and usability of \jjodel{}. Node.js introduced a lightweight event-driven back-end optimized for real-time interactions, providing the performance and responsiveness necessary for a collaborative modeling platform. React’s component-based architecture (not to be confused with that of the EMF ecosystem)  enabled modularity and reusability, allowing dynamic syntax customization and intuitive user interface design. However, adopting these frameworks was not straightforward; it required a significant paradigm change in application development and iterative testing to align their architectural models with the goals of \jjodel{}.

The broader evolution of the full-stack web ecosystem underscores the growing influence of these technologies. Key advancements such as Node.js (2009), Angular.js (2010), React (2013), and Vue.js (2014) have transformed the way modern applications are designed and developed. By 2024, Node.js powers over 40\% of web applications, cementing its role as a preferred backend technology in low-code platforms. React, used by 39.5\% of front-end developers, integrates seamlessly with advanced interfaces, making it a leading choice for creating responsive, component-based UIs\footnote{\url{https://www.statista.com/statistics/1124699/worldwide-developer-survey-most-used-frameworks-web/?utm_source=chatgpt.com}, last accessed on January 9, 2025}. These frameworks highlight the stark contrast in user experience between traditional modeling tools (all first released before), often emphasizing technical rigor, and modern low-code development platforms (LCDPs)~\cite{sahay2020supporting,prinz2021low}, which prioritize usability, real-time collaboration, and seamless adaptability.

For \jjodel{}, the decision to adopt Node.js and React was driven by the need to balance accessibility, flexibility, and \textit{richness}—qualities often absent in traditional tools. Node.js provided the foundation for real-time interactions, while React’s modular architecture empowered the platform to deliver an intuitive and customizable user experience. Together, these technologies helped \jjodel{} align its design with the broader goals of scalability and usability, ensuring that the platform could meet the diverse needs of its users. This process underscored the importance of carefully integrating technological frameworks to shape not only the technical implementation but also the overall user experience.

Crucially, these insights were not evident at the beginning. They emerged through a process of trial and error, with one particularly challenging iteration spanning two years of development. This iterative approach, guided by the expertise and advocacy of one of the authors, ultimately enabled the team to make informed decisions and implement a modern, scalable architecture for \jjodel{}.

In summary, the development of \jjodel{} reinforced a critical realization: technology is not neutral. It defines both the possibilities and constraints of a platform, influencing its ability to innovate and its inherent limitations. 

\subsection{Balancing Flexibility and Simplicity}
Balancing the flexibility of the platform with its usability proved to be one of the most significant challenges in the development of \jjodel{}. Many features were essential to meeting the requirements we add in mind~\cite{di2023jjodel}, but their integration risked introducing unnecessary complexity. To address this, \jjodel{} adopted a progressive disclosure approach, where essential features were initially exposed and advanced functionalities could be revealed later. Like this, both novice and experienced users could interact with the platform more effectively.

The decision to prioritize simplicity without compromising flexibility was further informed by user feedback. Iterative testing demonstrated that users, particularly those new to MDE tools, found overwhelming interfaces to be a barrier to productivity. By simplifying initial interactions and allowing users to explore more advanced features, \jjodel{} demonstrated that even complex modeling platforms could cater to diverse audiences while retaining powerful capabilities. 

\subsection{Avoiding Accidental Complexity}

Reducing accidental complexity~\cite{atkinson2008reducing} was a cornerstone of the \jjodel{} project. Traditional MDE tools often require complex installation procedures and specialized technical expertise, which hinder accessibility for students and domain experts. By prioritizing a client-side intelligence model, \jjodel{} eliminated many of these barriers, enabling users to engage with the platform without requiring advanced setup or configuration.

This decision significantly improved the accessibility and usability of the platform, aligning with its mission to democratize modeling practices. In addition, the streamlined architecture and processes reduced the cognitive load for users, allowing them to focus on their tasks rather than the intricacies of the tool itself. This approach highlights the importance of designing platforms that prioritize usability without sacrificing flexibility or power.

\subsection{Rethinking Concrete Syntax}

The integration of React’s component-based architecture into \jjodel{} fundamentally transformed how concrete syntax was conceptualized and rendered. Unlike traditional modeling tools, which often rely on static and rigid representations, the proposed approach allowed dynamic customization and real-time interaction. This flexibility enabled users to define syntax viewpoints tailored to their specific needs, fostering a more intuitive modeling experience.

This innovation also demonstrated the potential of modern web technologies to drive advancements in traditionally static domains. Using React's state management and rendering capabilities, \jjodel{} bridged the gap between abstract modeling concepts and practical applications, providing users with a seamless and interactive experience.

\subsection{Iterative Design and Feedback}

The iterative development process~\cite{4668134} was central to the success of \jjodel{}. Regular testing with novice and experienced modelers revealed critical gaps in usability and functionality, informing refinements in interface design and platform features. Early feedback emphasized the importance of real-time feedback mechanisms and context-aware interfaces, leading to the implementation of progressive disclosure and other user-centric design principles.

This approach ensured that \jjodel{} evolved as a responsive and adaptive platform, capable of meeting the diverse needs of its user base. The iterative process also reinforced the value of engaging with stakeholders throughout development, ensuring that the platform remained aligned with user expectations and requirements.

\subsection{Navigating Trade-offs}

Departing from established standards, such as Eclipse GLSP~\cite{bork2023vision}, presented significant challenges and unique opportunities for \jjodel{}. This decision enabled the platform to align more closely with its meta-metamodel, a strict extension of Ecore, and to prioritize client-side intelligence. IAlthough it resulted in a less standardized architecture, the trade-off proved advantageous, making the platform more reactive and agile while eliminating cumbersome integration processes often associated with traditional frameworks.

This experience highlights the critical importance of carefully evaluating trade-offs in platform design. By prioritizing flexibility and usability over strict adherence to established frameworks, we took a calculated risk that ultimately paid off. The resulting architecture demonstrated that divergence from canonical approaches, when guided by clear objectives, can lead to innovative and effective solutions aligned with the vision and goals of the platform.

\subsection{Exploring Alternatives to OCL.js}
Our experience with OCL.js\footnote{\url{https://ocl.stekoe.de/}}, a JavaScript-based implementation of the Object Constraint Language (OCL), provided both opportunities and challenges during the development of \jjodel{}. Despite OCL.js offered full integration with \jjodel{} and initially appeared to be a natural fit, its limitations became apparent over time. The lack of active development of the project and its partial coverage of the OCL standard forced us to explore alternatives to define constraints and predicates within the platform.

The most natural alternative was to adopt the declarative and functional expressions of JSX, which \jjodel{} already uses to template. This approach allowed us to leverage the same notation for both templating and predicates, creating a consistent and simplified user experience. However, it also introduced new uncertainties. To the best of our knowledge, JSX has not been formally characterized and no direct comparison between JSX and OCL has been documented in the literature. This absence of formalization leaves open questions about the expressive power, correctness, and potential limitations of JSX as a substitute for OCL.

This experience underscores the importance of evaluating the trade-offs between adopting emerging notation like JSX and relying on established but limited standards like OCL.js. Although JSX has served \jjodel{} well in the short term, further research and validation is necessary to determine its suitability as a foundational element to define constraints and predicates.

\subsection{The Value of Learning Through Failure}

Some of the most valuable lessons emerged from the failures encountered during development. Iterative attempts to align the architectural design with the objectives of the platform revealed the complexity of creating a scalable, reactive system~\cite{lyytinen1999learning}. In one instance, a two-year iteration led to significant revisions, demonstrating the importance of persistence and openness in addressing setbacks.

This experience reinforced the value of accepting failure as an opportunity for growth. Learning from missteps, \jjodel{} was able to refine its approach and develop a platform that is both innovative and user-friendly. The development of \jjodel{} underscored the importance of balancing technical innovation with user-centric design. The lessons learned highlight the need for thoughtful technology choices, iterative development, and a willingness to embrace trade-offs and failures as opportunities for growth. These insights provide a foundation for advancing the field of MDE tools, ensuring that future platforms are both powerful and accessible.

\section{Related Work}
\label{sec:rw}
This section reviews various MDE platforms, discussing their features, strengths, and limitations to position \jjodel{} within the current landscape of MDE tools.

MetaEdit+ \cite{Kelly2013} is a mature and widely recognized platform for domain-specific modeling (DSM). It uses the GOPPRR meta-metamodeling language \cite{kelly2005domain} to define domain-specific concepts and relationships, offering extensive customization capabilities. Despite its introduction in 1995 and its reliance on an outdated technology stack, MetaEdit+ features a reflective and integrated architecture that provides robust operational support through built-in governance mechanisms. These capabilities simplify complex tasks, such as ensuring metamodel consistency and managing artifacts, challenges that remain prevalent in many other MDE tools.
In comparison, \jjodel{} inherits the strengths of MetaEdit +, such as its reflective and integrated architecture. Leveraging a native cloud architecture and cutting-edge front-end technologies, \jjodel{} offers a more \textit{transparent} user experience by minimizing the accidental complexity, alongside support for real-time collaboration and automated syntax co-evolution. 

GMF/Eugenia\footnote{\url{https://eclipse.dev/modeling/gmp/}}, built on the Eclipse Modeling Framework (EMF), facilitates the automated generation of graphical editors. These tools are powerful in defining graphical representations. Unfortunately, their dependence on intricate configurations and technical complexity poses significant challenges for non-expert users. Moreover, their generative approach and component-based architecture hinder their ability to support dynamic adaptability, an essential feature for iterative and agile development workflows, as well as real-time collaboration. In contrast, \jjodel{} overcomes these limitations with an intuitive low-code environment that combines dynamic syntax customization with seamless collaboration, making it a more practical and efficient choice for modern team-based modeling scenarios.

Sirius\footnote{\url{https://eclipse.dev/sirius/}}, also built on the Eclipse ecosystem, introduces viewpoint-based modeling, allowing users to define and work with multiple perspectives in a single model. While this approach reduces programming effort, Sirius remains configuration-intensive and lacks native support for real-time collaboration or dynamic syntax co-evolution, limiting its adaptability in iterative and distributed development workflows. In contrast, the reflective architecture of \jjodel{} incorporates built-in governance mechanisms, often absent in tools such as Sirius, that simplify complex tasks such as coevolution and validation of metamodels, providing a more streamlined and robust modeling experience.

Building on the concepts of its predecessor, Sirius Web \cite{Giraudet2024} transitions to a native web environment, using React to deliver a modern interface. It introduces collaborative modeling features and supports various representation types, including diagrams, forms, and Gantt charts. While these advancements address some limitations of Sirius Desktop, Sirius Web still suffers from significant configuration overhead and limited dynamic adaptability, which constrain its effectiveness in highly iterative and fast-paced projects. Furthermore, its component-based architecture, rooted in EMF, lacks the seamless integration of an all-in-one environment, leading to fragmentation and increased complexity for users. 

Sprotty\footnote{\url{https://sprotty.org/}} and Gentleman \cite{3417990.3421998} address narrower roles in the MDE ecosystem. Sprotty is a web-based visualization framework focused on rendering lightweight models, while Gentleman provides a flexible projectional editing interface for directly modifying abstract syntax trees. Although these tools excel in specific use cases, they lack deeper integration with metamodel evolution and collaboration workflows. \jjodel{} combines the simplicity and lightweight design of tools like Sprotty and Gentleman with advanced capabilities such as automated syntax co-evolution, making it a more comprehensive solution.

A recent paper by Metin et al.~\cite{Metin2025} explores the challenges and lessons learned from working with GLSP and developing several GLSP-based modeling tools. The authors present a reusable reference architecture designed to facilitate the development and operation of GLSP-based web modeling tools. The reference architecture is then instantiated using BigUML as a running example. The paper highlights the procedural approach, key lessons learned, and critical reflections on the challenges and opportunities associated with using GLSP. In contrast to \jjodel{}, which natively supports cloud-based modeling, GLSP-based web modeling tools require the ad hoc implementation of the domain language, adding an additional layer of complexity.

HyperGraphOS \cite{Ceravola2024} represents an ambitious vision, integrating DSLs, modeling, and task execution into a unified system. Its infinite workspace concept and AI-enhanced adaptability demonstrate significant innovation, but these features come at the expense of usability for less technical users. 

Many of the platforms surveyed possess industrial-grade or commercial-technology readiness levels (TRL), making them well suited for high-stakes, large-scale projects. In contrast, \jjodel{} remains an academic endeavor, designed to explore and adapt the principles and solutions commonly found in large-budget projects and low-code platforms. Translating these into accessible and innovative approaches, \jjodel{} makes a unique contribution to the field. However, this academic focus also highlights its limitations in directly competing with fully commercialized solutions, particularly in terms of scalability, robustness, and enterprise-level support.
However, the initial adoption of \jjodel{} in academic settings has yielded promising results. Although a systematic evaluation of its impact on teaching and learning outcomes has not yet been conducted, preliminary feedback has been highly encouraging. Students have demonstrated significant engagement and understanding, even in the absence of extensive learning resources at the time of the course\footnote{For more details on the survey results, see \url{https://www.jjodel.io/student-survey/}.}. These early findings indicate that \jjodel{} has the potential to become a valuable educational tool, particularly in model-driven engineering courses.

\section{Conclusions and Future Work}
\label{sec:conclusion}
In this article, we present \jjodel{}, a cloud-based reflective modeling platform designed to address the challenges of cognitive complexity and usability of model-driven engineering (MDE). By emphasizing a low-code, user-friendly approach and integrating advanced features such as syntax viewpoints, real-time validation, and collaborative modeling, \jjodel{} bridges the gap between conceptual research and practical application in MDE. Through a detailed case study and a comprehensive analysis of its architecture and functionalities, we demonstrate how \jjodel{} empowers educators, professionals, and researchers to efficiently create, refine, and adapt domain-specific languages and models.

A core objective of \jjodel{} is to simplify the teaching of MDE concepts by providing an interactive and intuitive environment. Instructors often face significant challenges in adopting existing tools, which are frequently based on outdated technology stacks, and struggle to align them with modern teaching practices. By integrating familiar front-end technologies, such as JSX templates, \jjodel{} lowers the entry barrier, making it more accessible to students and allowing MDE to be effectively introduced at the undergraduate level. In addition, the platform supports for agile workflows and its ability to adapt to dynamic requirements make it an excellent choice for developing robust and scalable modeling solutions in industrial contexts.

Key innovations, including progressive disclosure, modular syntax viewpoints, and real-time collaboration, establish \jjodel{} as a new benchmark for MDE tools. These features reduce cognitive barriers and improve flexibility and scalability, making \jjodel{} a versatile tool for a wide range of use cases.

The development of \jjodel{} provided critical insight into the trade-offs and challenges inherent in the building of modern MDE tools. The implementation of advanced features, such as real-time collaboration and undo/redo mechanisms, required sophisticated state management and synchronization protocols. Furthermore, the adoption of React’s unidirectional data flow architecture enabled powerful functionalities like semantic zooming and dynamic syntax customization, but also presented scalability challenges when applied to large-scale models. These experiences also identified the importance of interdisciplinary approaches, such as taking advantage of insights from related fields like online gaming, to tackle complex design challenges.

Looking ahead, \jjodel{}’s development will focus on addressing scalability for industry-scale models by adopting advanced optimization techniques from the broader communities of MDE and distributed systems. Key areas for future work include the following challenges:
\begin{itemize}
    \item Enhancing rendering and synchronization mechanisms to efficiently support larger and more complex models.
    \item Advancing automation for metamodel and model co-evolution to accommodate a broader range of use cases and minimize manual intervention.
    \item Developing sophisticated conflict resolution mechanisms and version control systems tailored for distributed teams.
    \item Exploring integration with existing MDE platforms and tools, such as the Epsilon Playground\footnote{\url{https://eclipse.dev/epsilon/playground/}}, to enable functionalities such as the execution of model-to-model transformations directly within the environment of \jjodel{}.
    \item Consolidating ongoing experiments with large language model (LLM) agents to further enhance modeling capabilities;
    \item Expanding the \jjodel{} community by releasing open-source components, encouraging contributions, and establishing an organization dedicated to securing funding and ensuring the project’s sustainability over the long term.
\end{itemize}

By continuing to evolve and address these challenges, \jjodel{} aims to establish itself as a cornerstone tool for MDE, serving both as a practical resource for immediate application and as a foundation for innovation in the field. Its commitment to reducing complexity, fostering collaboration, and supporting adaptability ensures that \jjodel{} will remain a relevant and impactful tool as MDE demands continue to grow.

\subsection{A Roadmap for the Future}

The lessons learned during the development of \jjodel{} provide a roadmap for the future of MDE platforms. Reducing cognitive complexity, fostering collaboration, and supporting adaptability are essential to ensure widespread adoption of MDE practices. These principles make \jjodel{} a valuable resource for practitioners today and a foundation for future innovation.  

However, sustaining such advances requires more than technical expertise: it also demands robust community support and funding. Unlike in the early 2000s, when national and international research programs funded tool development, today’s funding landscape is less aligned with such objectives. This shift raises important questions about the responsibility for developing and maintaining high-quality modeling platforms.  

Although opinions within the MDE community vary, we advocate a shared commitment to advancing both foundational theories and practical solutions. High-quality tools are essential to demonstrate the effectiveness of MDE and ensure its continued relevance in various domains. By working together, the MDE community can build platforms that empower users, enhance education, and drive innovation across industries.

\bibliographystyle{abbrv}  
\bibliography{bib.bib}

\end{document}